\documentclass[useAMS,usenatbib]{mn2e}
\usepackage{aas_macros}
\usepackage{graphicx}
\usepackage{hyperref}
\usepackage{amsmath,bm}

\newcommand{\rr}{$r_V$}
\newcommand{\mrr}{r_V}

\newcommand{\rv}{$R_V$}
\newcommand{\av}{$A_V$}

\newcommand{\mav}{A_V}
\newcommand{\avg}[1]{\left< #1 \right>} 

\newcommand\arcdeg{\mbox{$^\circ$}}%

\begin{document}

\title[Bayesian Extinction]{Evidence for Grain Growth in Molecular Clouds: A Bayesian Examination of the Extinction Law in Perseus}
\author[J. Foster, K. Mandel et al.]
{\parbox{\textwidth}{Jonathan B. Foster$^{1}$\thanks{E-mail: \texttt{jonathan.b.foster@yale.edu}}, 
Kaisey S. Mandel$^{2}$\thanks{E-mail: \texttt{k.mandel@imperial.ac.uk}},
Jaime E. Pineda$^{3}$,
Kevin R. Covey$^{4}$,
H\'{e}ctor G. Arce$^{5}$,
Alyssa A. Goodman$^{6}$}\vspace{0.4cm}\\
\parbox{\textwidth}{
$^{1}$Institute for Astrophysical Research, Boston University, Boston, MA 02215, USA. Current Address: Department of Astronomy Yale University, P.O. Box 208101 New Haven, CT 06520-8101, USA.\\
$^{2}$Imperial College London, Blackett Laboratory, Prince Consort Road, London SW7 2AZ, UK \\
$^{3}$Current address: ESO, Karl Schwarzschild Str. 2, 85748 Garching bei Munchen, Germany; and UK ALMA Regional Centre Node, Jodrell Bank Centre for Astrophysics, School of Physics and Astronomy, University of Manchester, Manchester, M13 9PL, UK \\
$^{4}$Department of Astronomy, Cornell University, 226 Space Sciences Building, Ithaca, NY 14853, USA \\
$^{5}$Department of Astronomy, Yale University, P.O. Box 208101, New Haven CT 06520, USA \\
$^{6}$Harvard-Smithsonian Center for Astrophysics, 60 Garden Street, Cambridge, MA 02138, USA}}

\date{Accepted by MNRAS 2012 October 2}

\pagerange{\pageref{firstpage}--\pageref{lastpage}} \pubyear{2012}

\maketitle

\label{firstpage}

\begin{abstract}
We investigate the shape of the extinction law in two 1\arcdeg\ square fields of the Perseus Molecular Cloud complex. We combine deep red-optical ($r$, $i$, and $z$-band) observations obtained using Megacam on the MMT with UKIDSS near-infrared ($J$, $H$, and $K$-band) data to measure the colours of background stars. We develop a new hierarchical Bayesian statistical model, including measurement error, intrinsic colour variation, spectral type, and dust reddening, to simultaneously infer parameters for individual stars and characteristics of the population.   We implement an efficient MCMC algorithm utilising generalised Gibbs sampling to compute coherent probabilistic inferences.   We find a strong correlation between the extinction (\av) and the slope of the extinction law (parameterized by \rv). Because the majority of the extinction toward our stars comes from the Perseus molecular cloud, we interpret this correlation as evidence of grain growth at moderate optical depths. The extinction law changes from the ``diffuse'' value of \rv\ $\sim$ 3 to the ``dense cloud'' value of \rv $\sim$ 5 as the column density rises from \av\ = 2 mags to \av\ = 10 mags. This relationship is similar for the two regions in our study, despite their different physical conditions, suggesting that dust grain growth is a fairly universal process. \end{abstract}


\section{Introduction}

Interstellar dust affects virtually all astronomical observations, yet remains poorly understood. A full prescriptive theory which quantitatively explains the production, destruction and steady state conditions of interstellar dust does not exist although various simple models \citep[e.g.][]{Inoue:2011} have been proposed. In particular, the specific minerals, the shape (or structure), mass/size distribution, and total amount of interstellar dust cannot generally be produced from first principles. Much progress, though, has been made toward a descriptive theory. A basic model with separate power-law size distributions of bare graphite and silicate grains was shown to adequately reproduce most of the observational evidence by \citet{Draine:1984}.  \citet{Weingartner:2001} updated the model to include very small grains---polycyclic aromatic hydrocarbons (PAHs)---which are necessary both to reproduce observed mid-infrared spectral features and to explain the transiently heated grains identified by \citet{Sellgren:1984}. Although the \citet{Weingartner:2001} model is relatively simplistic (the model uses spheroidal grains although grains are probably ``fluffy'' and fractal) and slightly ad hoc (the silicate is ``astronomical amorphous silicate'' with a few lab-measured spectral features modified to accommodate the observations), this model has proven to be quite robust and is still widely used to interpret the extinction and emission features of dust from the X-ray through the mid-infrared \citep[see][for a review of this model's successes and shortcomings]{Draine:2003}.

Another nice feature of the \citet{Weingartner:2001} model is that it directly relates an observational parameter to an underlying physical change. In particular, the highly influential work of \citet{Cardelli:1989} (CCM) found that different extinction curves towards different lines of sight could all be adequately fit by a one-parameter family of curves:
\begin{equation}
A_{\lambda} = A_V (a_{\lambda} + b_{\lambda} R_V^{-1})
\end{equation}
where $R_V$ does double-duty as both the slope of the extinction law at $V$ band relating the colour excess $E(B-V)$ to the extinction $A_V$
\begin{equation}
R_V = \frac{A_V}{E(B-V)}
\end{equation}
and as the parameter distinguishing different extinction laws. The coefficients  $a_{\lambda}$  and $b_{\lambda}$ are high-order polynomial functions of wavelength in the optical, but simple power laws (with index $\sim$ 1.61) in the near-infrared \citep[see also][]{Mathis:1990, Rieke:1985}. The model of \citet{Weingartner:2001} explains this result by observing that as grains grow in their model the value of $R_V$ grows (in the sense that larger average grain sizes have larger $R_V$) and  the extinction law at other wavelengths changes consistent with the parameterization of CCM. \rv = 3.1 is a typical value for the ``diffuse'' interstellar medium (ISM). Dust growth may either be due to the accretion of icy mantles or from coagulation of fluffy dust grains. Thus, a connection is established between observational changes to the extinction law and the physical properties of the underlying model.

Inside dense clouds, dust undergoes significant processing. The two most important processes are coagulation (grains sticking together and growing) and mantle-accretion of volatile elements. Some accretion of a volatile mantle onto refractory dust grains is presumed to occur in cold dense regions because observations of gas in molecular cores show severe depletion of several important molecules at high densities. This is particularly true for carbon-bearing species \citep[see][and references therein]{Bergin:2007}. Indeed, a theory of grains where bare silicate cores are covered by carbon-rich mantles can do a fair job of explaining the ISM dust more generally \citep{Greenberg:1999}. Additionally, the mid-infrared spectra of dense regions often shows features which are interpreted as ice mantles of volatiles. At the same time, coagulation must be important, because the extinction per hydrogen atom ($A_V/N(H)$) is observed to \emph{decrease} in dense regions. This is interpreted by \citet{Mathis:1990} as requiring the coagulation of grains, which can be thought of as preventing the interiors of grains from effectively participating in extinction. Adding more material to the extinguishing dust via depletion without also coagulating would generally tend to \emph{increase} the total amount of extinction per hydrogen atom simply because there is more material blocking radiation.

Observations of background stars behind a column of dust provide the best way to directly measure the shape of the extinction law but the wavelength range must be chosen with care. Dust exhibits rich spectral features in the mid-infrared which are normally ascribed to PAHs. The strength of these features is expected to be a function of the precise chemical history of a population of dust grains and the particular radiation environment to which the dust is currently exposed. These features can also produce emission which contributes to the mid-infrared fluxes of background stars observed through a dust cloud, complicating the use of these wavebands and presumably explaining the disparate attempts to nail down this portion of the extinction curve \citep[compare][]{Indebetouw:2005, Flaherty:2007, Chapman:2009}. Wavelengths in the blue-optical or UV are quite sensitive probes of the extinction law, but it is prohibitively expensive to obtain observations of highly reddened stars at these wavelengths. The red-optical and near-infrared is thus the best place to study the question.

A single molecular cloud is a complex hierarchical structure, with a dense filamentary web and a vast range of physical environments. The cores that eventually form stars are often well-modelled as simple Bonner-Ebert spheres \citep[e.g.][]{Alves:2001,Schnee:2005}. Within such a centrally condensed spherical object the observed quantity---the total column density, which we shall call $A_V$ for conventional reasons---can be loosely related to the underlying physical driver of dust grain growth---the volume density---in the sense that increasing column density implies increasing volume density. Our chain of inference is thus \rv\ $\propto$ $\avg{a}$ $\propto$ $n$ $\propto$ \av, where $n$ is the volume density and $\avg{a}$ represents the average size of dust grains. Outside of a single centrally condensed object, the connection between column and volume density will break down. Nonetheless, because structures in molecular clouds tend to be centrally condensed this chain of inference may hold, at least for the portions of the cloud were we are predominantly seeing through a single structure. 

Previous studies of changes in extinction law in molecular clouds have mostly been at large \av\ (\av $>$ 10 magnitudes). \citet{Cambresy:2011} report a transition in the extinction law at \av\ = 20 magnitudes in \emph{Spitzer} IRAC bands (3.6, 4.5, and 5.8 \micron). \citet{Roman-Zuniga:2007} measured the extinction law at large optical depths (up to \av\ of 60 magnitudes) in B59 and found no evidence for variation in the extinction law at these depths; the favoured extinction law through the entire cloud was the \citet{Weingartner:2001} model with \rv\ = 5.5. \citet{Cambresy:2005} report a change in the extinction properties around \av\ = 1 magnitude averaged over the galactic anti-centre hemisphere. This analysis is based on a comparison against the \citet{Schlegel:1998} dust emission maps maps, so this could also correspond to the regime where the \citet{Schlegel:1998} dust temperature correction is uncertain \citep{Arce:1999b}. Studies using other methods have also inferred changes in dust properties as a function of increasing depth inside a cloud using, for example, scattered light \citep[e.g.][]{Steinacker:2010} or changes in the dust emissivity at submillimetre wavelengths \citep[e.g.][]{Stepnik:2003}.

In this study, we combine deep multi-wavelength data in the red-optical ($r,i,z$) with near-infrared ($J,H,K$) data from the UKIRT Infrared Deep Sky Survey (UKIDSS) for large sections of the Perseus molecular cloud in order to test the relationship between \rv\ and \av\ over the range of column densities where the diffuse ISM-like matter of the molecular cloud transitions into the dense-core regime.  We employ a novel hierarchical Bayesian approach to statistically model multiple sources of uncertainty and coherently estimate the parameters for individual stars and the population.  We find strong evidence that there is a global correlation in the sense that the extinction law becomes steeper (\rv\ becomes larger which corresponds to larger dust grains in the model of \citet{Weingartner:2001}) as \av\ increases from around 2 magnitudes to about 10 magnitudes of extinction.

In \S~\ref{sec:data} we describe the acquisition, reduction, and calibration of our photometric data. In \S~\ref{section:fitting} we describe our model to fit the photometric data using an empirical colour locus and a parameterized extinction law. We describe a traditional least-squares approach (\S~\ref{Least-Squares}) as well as a hierarchical Bayesian model (\S~\ref{section:bayesian}) to overcome some of the problems in the least-squares approach. We present the results from this hierarchical Bayesian model in \S~\ref{sec:results}, and we conclude in \S~\ref{sec:conclusion}. We describe the mathematical details of our statistical model in Appendix A. In Appendix B, we present details of our MCMC algorithm for computing statistical inferences.

\section{Data}
\label{sec:data}
\subsection{Megacam Data Acquisition}

\begin{figure*}
\includegraphics[width=16cm]{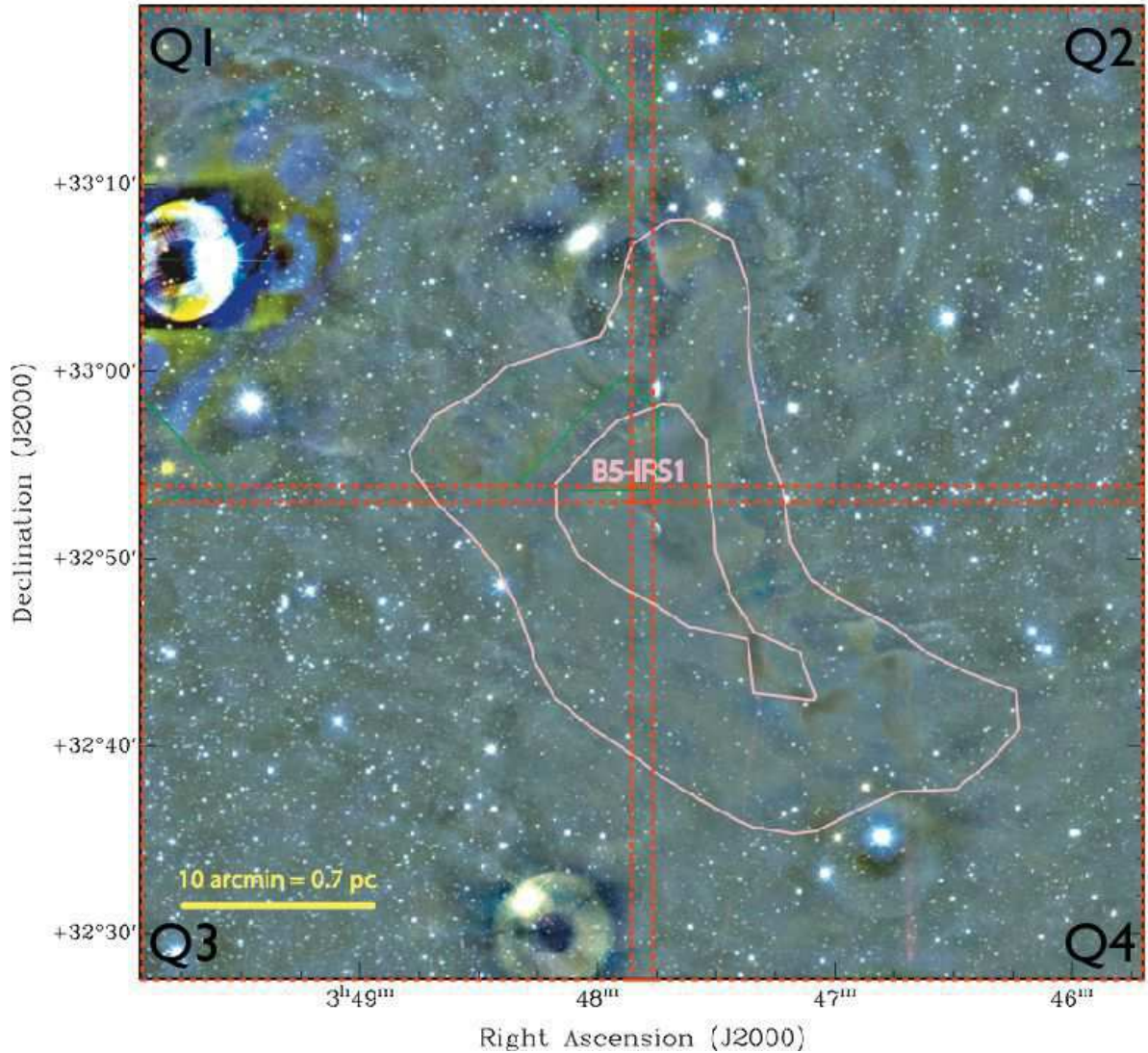}
\caption{A three-colour ($r,i,z$) image of B5 with contours at 3 and 5 mags of \av\ from a 2MASS-based extinction map \citep{Ridge:2006}. Also shown are the four quadrants used in observing. The most energetic young star, B5-IRS1 is visible at the centre with a small reflection nebula around it. Processing artefacts include the ``donut'' shaped ghosts around bright stars and streaks from satellites. In the top-left of the image, green lines show areas not covered by $r$ band due to a mistaken alignment. The bright star in this corner caused problems for the de-fringing algorithm, resulting in a corrupt image here.}
\label{B5-picture}
\end{figure*}

\begin{figure*}
\includegraphics[width=16cm]{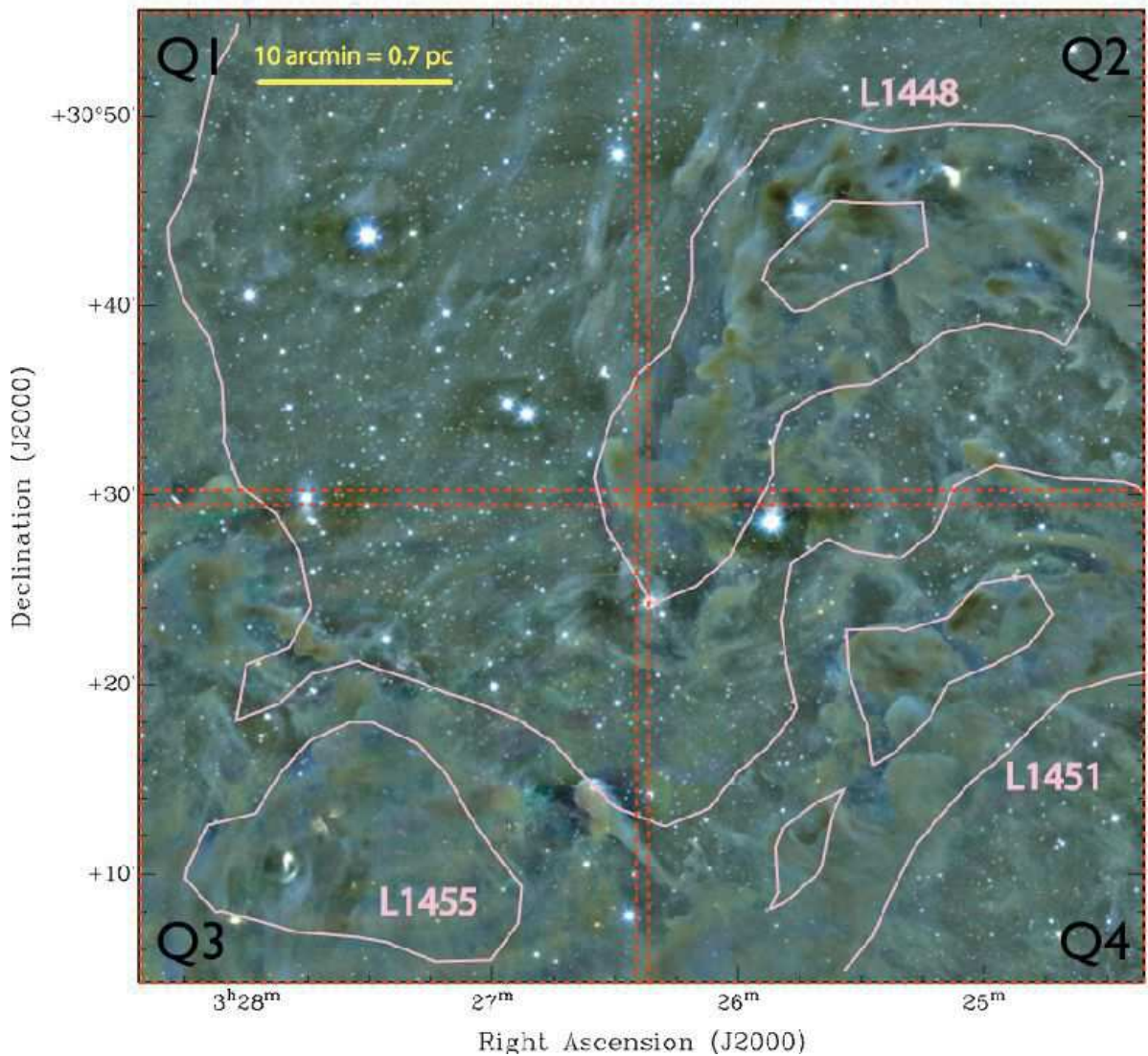}
\caption{A three-colour ($r,i,z$) image of our data from the West-End of Perseus with pink contours  at 3 and 5 mags of \av\ from a 2MASS-based extinction map \citep{Ridge:2006}. Also shown are the four quadrants used in observing. The three main clouds, L1448, L1451, and L1455 are labelled. L1448 and L1455 contain young stars. Visible remaining artefacts include the faint ``donut'' shaped ghosts around bright stars, streaks from some satellites or asteroids, and some residual fringing around the ``comet'' in the lower-centre of the image where the scaling algorithm failed.}
\label{WE-picture}
\end{figure*}

Our red-optical data were taken during the second halves of the night on November 2nd and 3rd, 2007 using Megacam at the 6.5-m MMT on Mt. Hopkins\footnote{Megacam has since moved to the Magellan Clay telescope at Las Campanas Observatory}. We observed two fields within the Perseus molecular cloud complex; one field was centred on B5 (R.A., Dec = 56.96$^{\circ}$, +32.84$^{\circ}$; Figure~\ref{B5-picture}) and the other field was centred on the ``West-End'' of Perseus \citep[see][for the definition of this region]{Pineda:2008}, containing L1448, L1451, and L1455 (R.A., Dec = 51.58$^{\circ}$, +30.50$^{\circ}$; Figure~\ref{WE-picture}). Individual exposures were 80 seconds in $r$, 60 seconds in $i$ and 50 seconds in $z$. The general pattern was to observe all of one field each night, so the West-End (WE) was observed on 11/2/07, while B5 was observed on 11/3/07. However, a few additional WE images were obtained on 11/3/07 at the end of the night. 

Science observations were interlaced with images of our calibration field in order to provide airmass solutions. Our calibration field was Per-Cal, a portion of the Perseus supercluster of galaxies included in the Sloan Digital Survey Extension for Galactic Understanding and Exploration \citep[SDSS-SEGUE;][]{Aihara:2011} and centred at R.A., Dec = (54.9755$^{\circ}$, +41.1533$^{\circ}$). All $r$-band science images and calibration images were taken first during the night, and were thus taken between an airmass of 1.0 and 1.06, providing insufficient lever arm to fit for an airmass correction term. Total exposure times were 800 seconds in $r$, 600 seconds in $i$ and 500 seconds in $z$. For $i$ and $z$, data were typically taken between 1.1 and 2.0 airmass and the observations were interleaved with 5 repetitions of the following pattern: $i$,$z$,$z$,$i$. Each field was observed as four different quadrants, with some overlap between quadrants (see Figures~\ref{B5-picture} and \ref{WE-picture}) and each quadrant was observed contiguously in $i$ and $z$. Thus, for an individual quadrant, the $r$ band images were taken near zenith and then the $i$ and $z$ observations were taken in an interleaved fashion later in the night.

The weather was generally good but not perfectly photometric. On 11/02/2007 the seeing was excellent with typical values of 0.6\arcsec, sometimes improving to 0.4\arcsec. Some cirrus clouds were observed during the first half of the night (before our data were collected) and were occasionally observed during the early part of our observing (mostly during calibration images of Per-Cal). On 11/03/2007 the seeing was relatively poor (0.7 - 1\arcsec). Again, a few clouds were observed earlier in the night, but then cleared. In B5, the $r$-band images for the first quadrant were taken at 45\arcdeg\ with respect to north, causing us to have some gaps in our final images and catalogue (see Figure~\ref{B5-picture}). 

\subsection{Megacam Data Reduction}

Data reduction of the Megacam data followed a largely standard set of procedures in IRAF \citep[Image Reduction and Analysis Facility;][]{Tody:1993}. Bias and flat frames were made and used to process the images. Bad pixels were identified and cosmic rays removed. Significant interference fringing (where unabsorbed light reflects off the bottom of the CCD and interferes with incoming radiation) is present in the $i$ and $z$ images, and the de-fringe program by B. McLeod was used to remove it. Megacam is made up of 36 individual CCDs. Because the de-fringing algorithm scales the intensity of the fringe pattern before subtracting it from an individual CCD, in places where individual bright stars or strong surface brightness features dominate the background for an entire individual CCD the fringes are poorly subtracted. Bright stars also cause some ``ghosts'' or halos on the final images, where light scatters internally in the telescope optics, and these increased surface brightness causes additional problems for the de-fringing scaling. Defects from bright stars and other features can be seen in both Figures~\ref{B5-picture} \& \ref{WE-picture}.

An illumination correction was also applied, although this correction was only significant at $r$. This correction removes most large variations in zero-point from one CCD chip to the next, but in practice this correction was not sufficient, and we applied a different zero-point for each CCD in the calibration.

We used {\sc swarp} \citep{Bertin:2002} to co-add individual images (weighting was based on measured noise within the images) to produce Figures~\ref{B5-picture} \& \ref{WE-picture}, and source detection was performed on these deep co-added images to use in cross-referencing the catalogues. Calibrated photometry was performed on each frame individually (so that the airmass and zero-point corrections discussed in the following section could be applied) and final source magnitudes were the noise-weighted average of the magnitudes measured in each frame.

\subsection{Megacam Data Calibration}

Because we wish to fit a parameterized main sequence in SDSS colours, it is necessary to place our Megacam photometry onto the SDSS ($r,i,z$) system. This process accounts for variations in filter and atmospheric transmission at different sites, as well as any other optical features of the telescope which may change the efficiency with which a flux at a given wavelength is measured. Literature colour corrections of this type are often based on only a small number of comparison standard stars, and are thus rather perilous. For example, \citet{Carpenter:2001} use roughly 50 stars to derive the widely-used transformation equations between Two Micron All Sky Survey \citep[2MASS;][]{Skrutskie:2006} and other near-infrared photometric systems. Our calibration fields contain thousands of stars in the SDSS catalogue and were taken concurrently with our data, reducing the influence of variable atmospheric transparency. This provides us with a robust way to calculate these transformations. 

For each filter, $x$, we solve for CCD chip-dependent photometric zero-points ($ZP_{x,j}$) and chip-independent airmass ($c_{x}$) and colour terms ($d_{x}$) by fitting for
\begin{eqnarray}
\mathrm{Inst}_{r} & = & \mathrm{SDSS}_{r} + ZP_{r,j} + d_{r} (\mathrm{SDSS}_{r} - \mathrm{SDSS}_{i}) \\
\mathrm{Inst}_{i} & = & \mathrm{SDSS}_{i} + ZP_{i,j} + c_{i} A + d_{i} (\mathrm{SDSS}_{r} - \mathrm{SDSS}_{i}) \\
\mathrm{Inst}_{z} & = & \mathrm{SDSS}_{z} + ZP_{z,j} + c_{z} A + d_{z} (\mathrm{SDSS}_{r} - \mathrm{SDSS}_{i})
\end{eqnarray}
where $A$ is the airmass at which a particular image was taken. Note that because all our $r$ band images were taken at low airmass, we had no leverage on the airmass term in this filter and thus do not fit for it. However, since our data $r$ band images were also taken at the same low airmass, the lack of this term will not significantly influence our calibration. We solved for the calibration parameters for each night separately to account for the different conditions, and list the resulting fits in Table~\ref{Calibration}.

\begin{table}
\begin{tabular}{lccc}\hline
Filter (x) & Zero-Points (ZP$_{x,j}$) & Airmass (c$_x$) & Colour (d$_x$) \\\hline
\multicolumn{4}{c}{WE} \\
      $r$ &    -1.01 -- -1.08   &  N/A & -0.127 \\
      $i$  &   -0.28 -- -0.32  & 0.0104 & -0.153 \\
      $z$ &    0.8 -- 0.9  &  0.073 & -0.0089  \\
\multicolumn{4}{c}{B5} \\
      $r$  &   -1.0 -- -1.1  &   N/A & -0.133  \\
      $i$  &    -0.21 -- -0.30 & 0.033 &  -0.156  \\
      $z$ &   0.9 -- 1.0 & 0.093 & -0.0067 \\ \hline
\end{tabular}
\caption{\label{Calibration} Calibration to SDSS. Colour term is SDSS$_r$ - SDSS$_i$ for $r$, $i$, $z$. See Eqns. 3-5 for the definition of these coefficients.}
\end{table}

\begin{figure*}
  \begin{center}
    \begin{tabular}{cc}
      \resizebox{8cm}{!}{\includegraphics[angle=0]{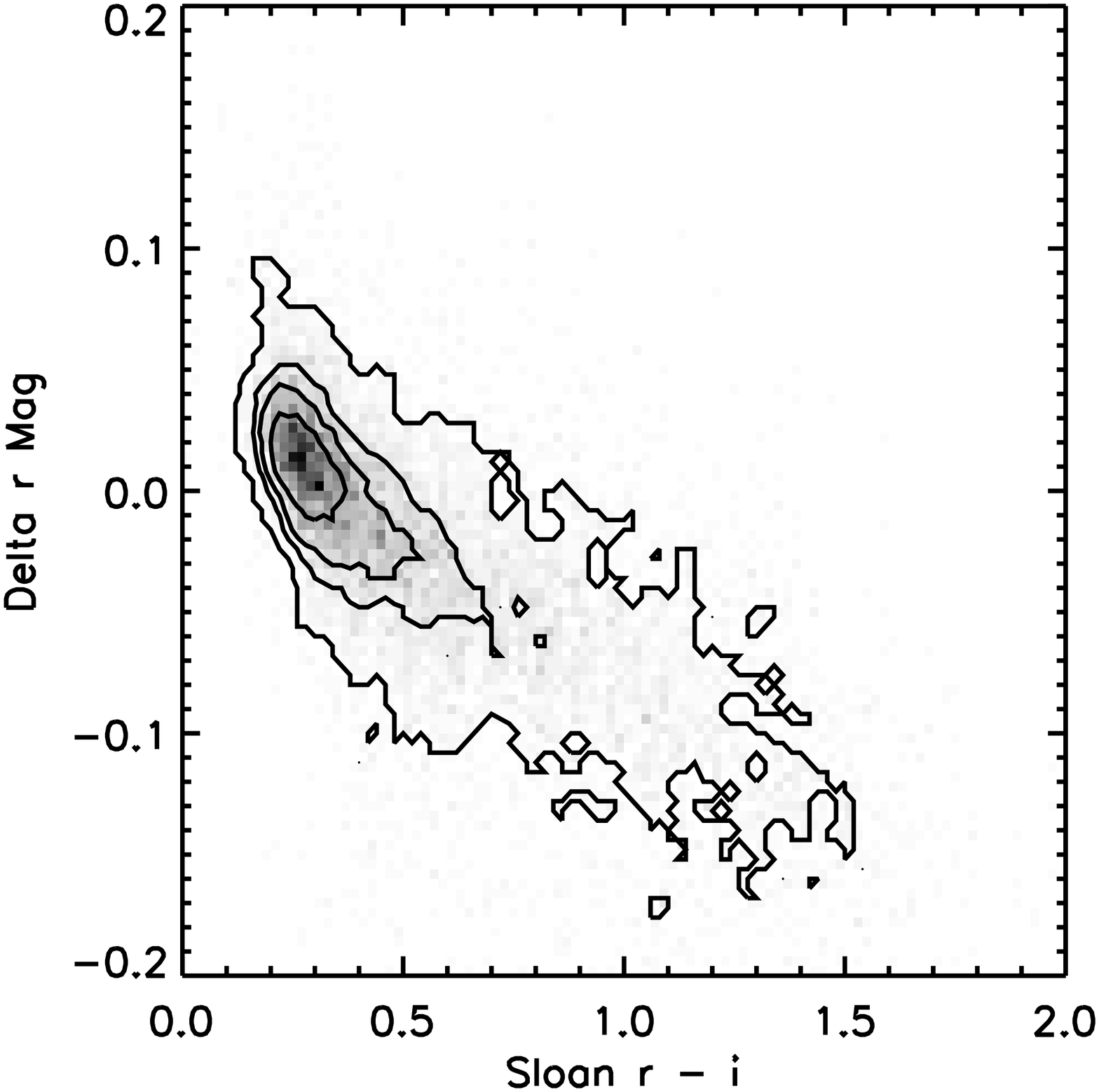}}&
     \resizebox{8cm}{!}{\includegraphics[angle=0]{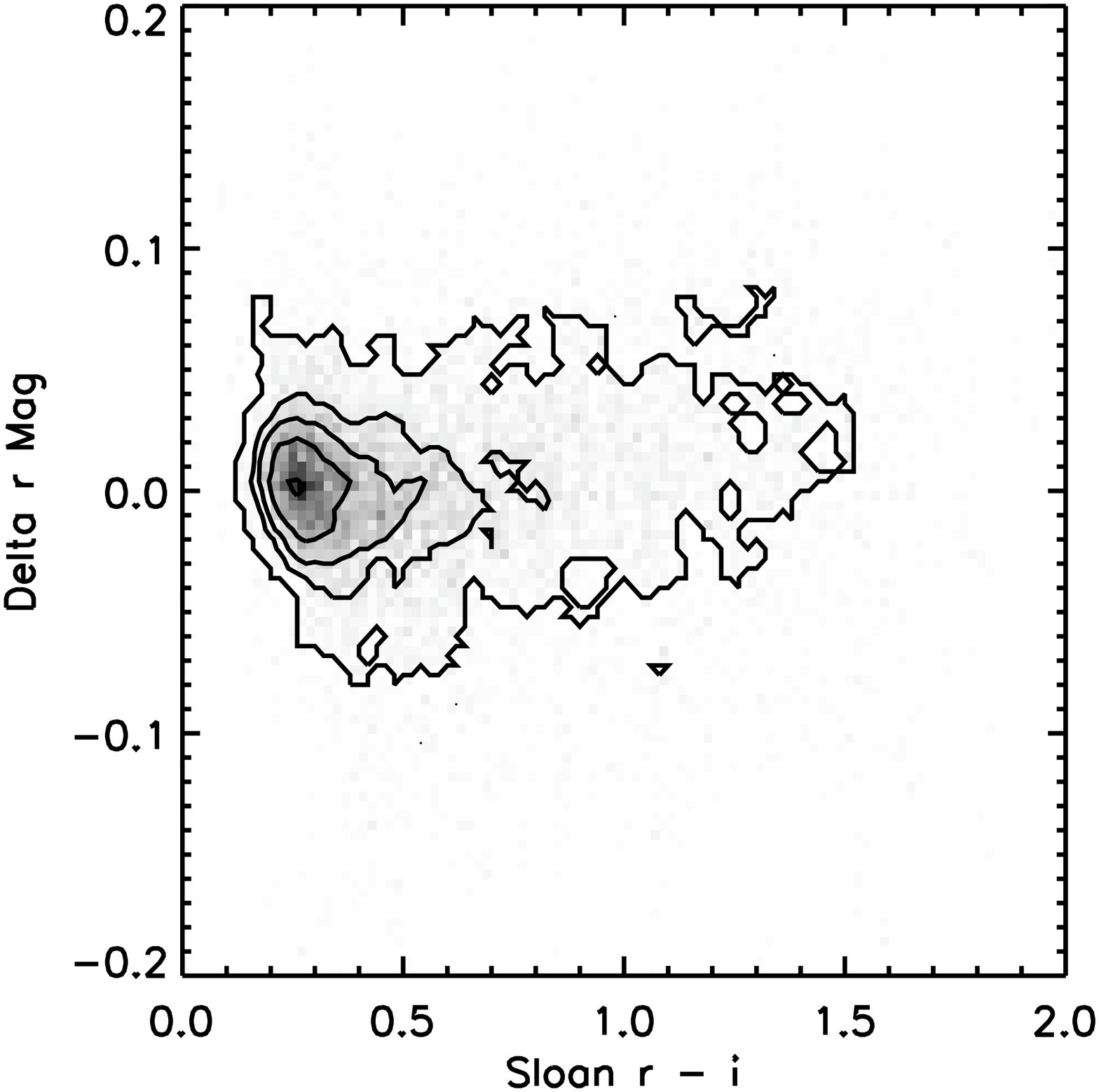}}
    \end{tabular}
\caption{The effect of colour correction in $r$ band. Left-panel shows (MMT - SDSS) $r$-band magnitude after solving for zero-point but before correcting for the colour term. Right-panel shows the same difference after applying our colour term.}
\label{rCalibration}
  \end{center}
\end{figure*}

\begin{figure*}
  \begin{center}
    \begin{tabular}{cc}
      \resizebox{8cm}{!}{\includegraphics[angle=0]{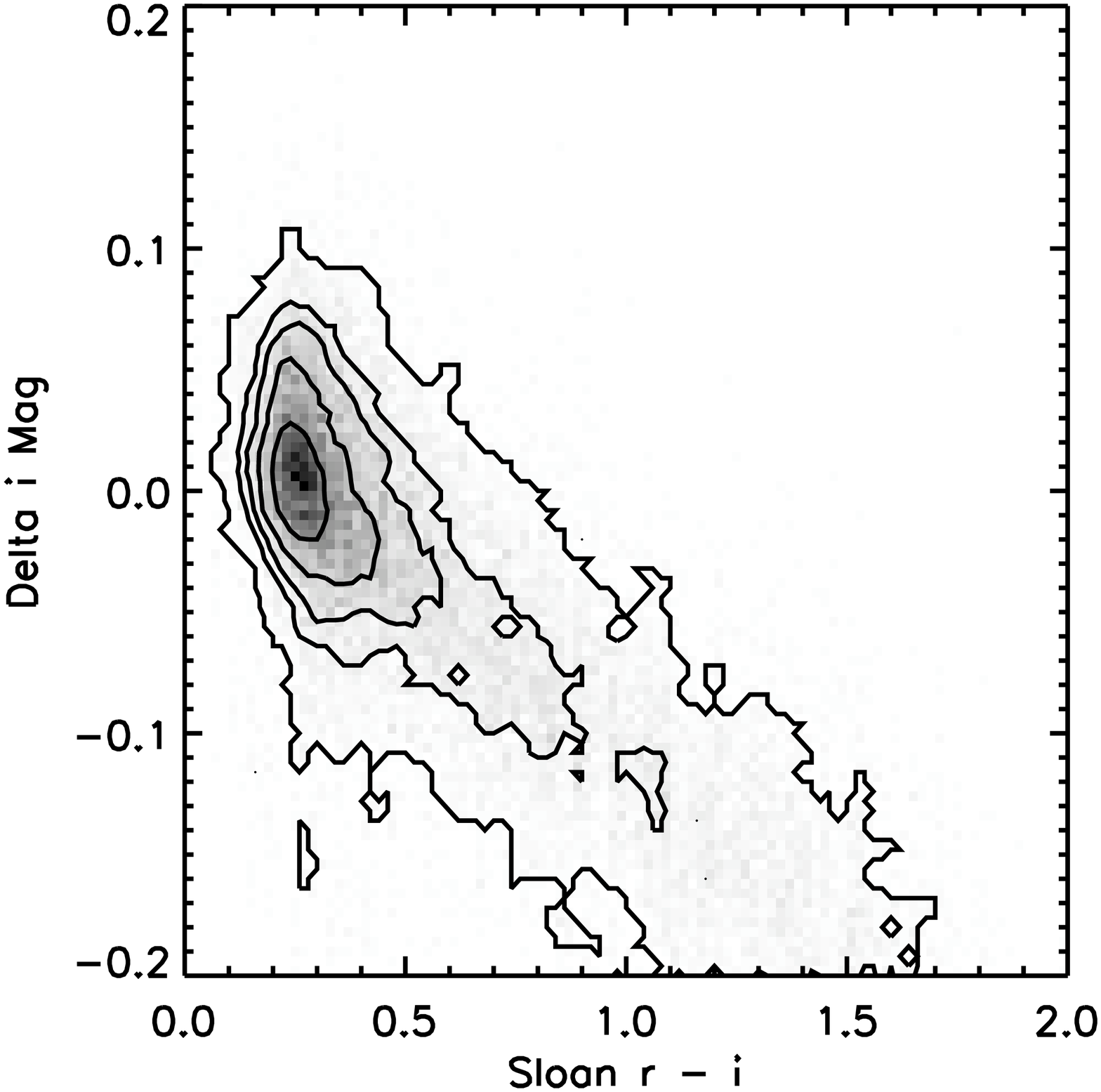}}&
     \resizebox{8cm}{!}{\includegraphics[angle=0]{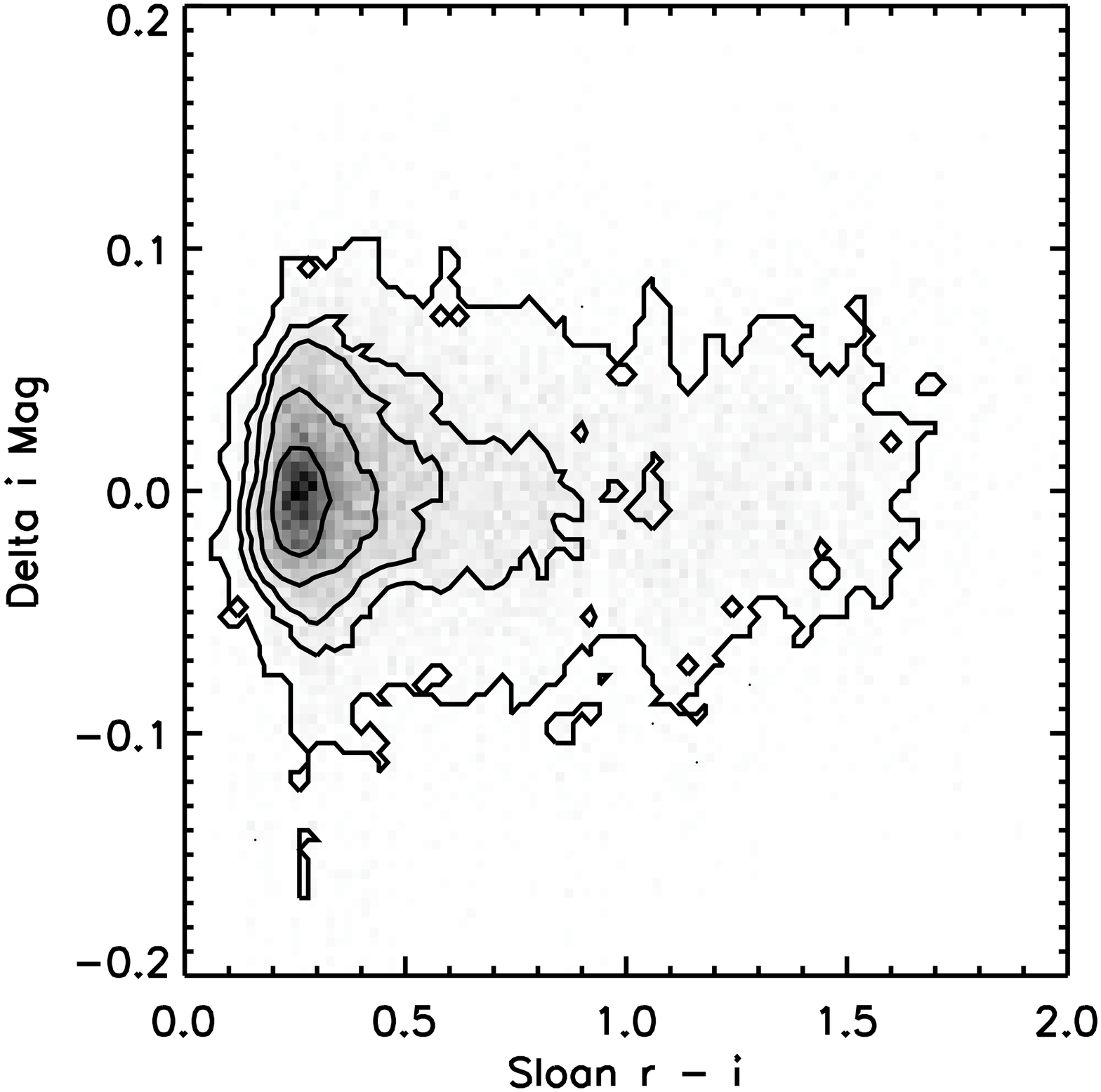}}
    \end{tabular}
\caption{The effect of colour correction in $i$ band. Left-panel shows (MMT - SDSS) $i$-band magnitude after solving for zero-point but before correcting for the colour term. Right-panel shows the same difference after applying our colour term.}
\label{iCalibration}
  \end{center}
\end{figure*}

\begin{figure*}
  \begin{center}
    \begin{tabular}{cc}
      \resizebox{8cm}{!}{\includegraphics[angle=0]{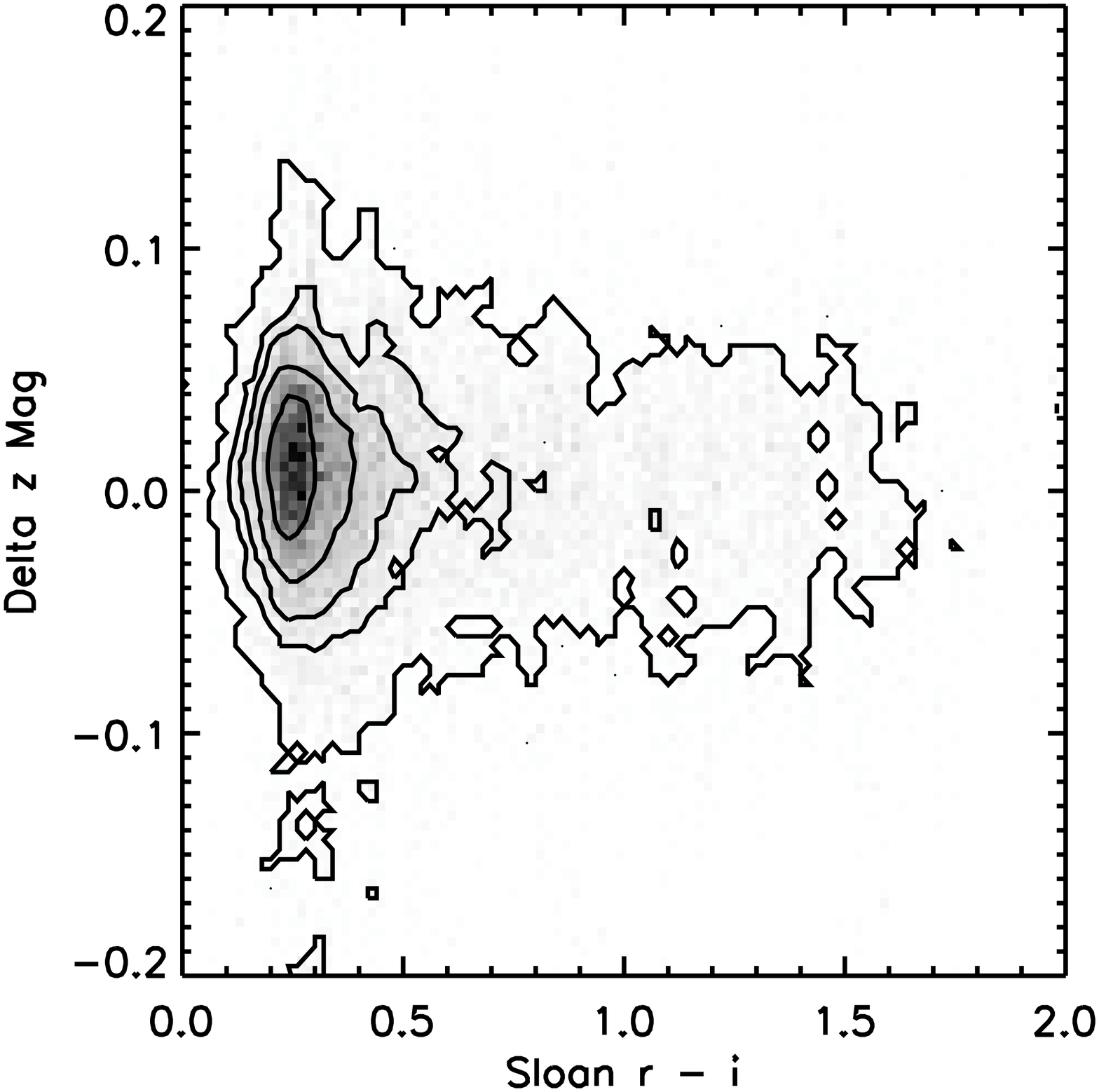}}&
     \resizebox{8cm}{!}{\includegraphics[angle=0]{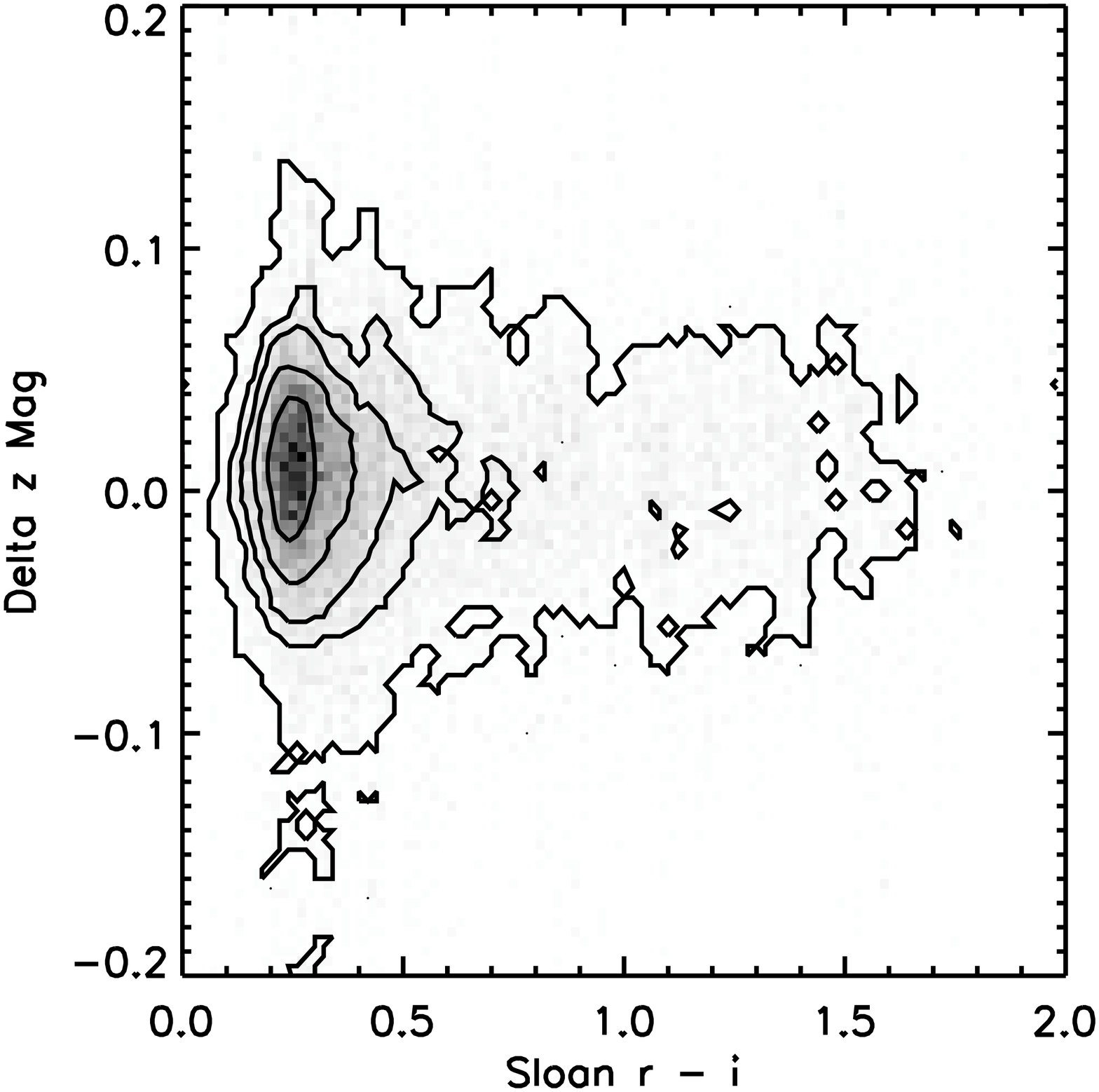}}
    \end{tabular}
\caption{The effect of colour correction in $z$ band. Left-panel shows (MMT - SDSS) $z$-band magnitude after solving for zero-point but before correcting for the colour term. Right-panel shows the same difference after applying our colour term. Note that the correction was insignificant for this band.}
\label{zCalibration}
  \end{center}
\end{figure*}

The colour terms ($d_{x}$) proved to be quite significant for the $r$ and $i$ bands but insignificant for $z$. We show the results of applying this calibration to the control field stars for the second night in Figures~\ref{rCalibration}--\ref{zCalibration} for $r$, $i$, and $z$ respectively. Figures for the first night of data are similar.  We experimented with expanding the colour correction to include a term proportional to $\mathrm{SDSS}_{i} - \mathrm{SDSS}_{z}$ but found that this additional complexity did not significantly reduce the observed scatter around the calibration relation. Plots similar to Figures~\ref{rCalibration}--\ref{zCalibration} but versus $i$-$z$ showed that applying the $r-i$ colour correction corrected this colour as well.

Preliminary results of this analysis were presented by \citet{Foster:2008}, combining these Megacam observations with 2MASS data to infer a relationship between \av\ and \rv\ similar to what we report here. The photometric uncertainties of the 2MASS data, however, dominated the error budget of this prior analysis and motivated us to modify our analysis to incorporate higher quality NIR photometry from the UKIDSS survey (see \S~\ref{ukidss}).  The subsequent improvement in our error budget also revealed the presence of a photometric offset in the fourth quadrant of the B5 MegaCam photometry. 

Because the four quadrants for each field overlapped slightly, we were able to check the reliability of our calibration by comparing the calibrated magnitudes for stars appearing in multiple quadrants. The spread in magnitude was typically consistent with our estimated measurement and calibration uncertainty. The only exception was the fourth quadrant in B5. For this quadrant, magnitudes were not consistent with the other three quadrants (the other three were all consistent). We found the median offset in the overlapping stars and offset the magnitudes in the fourth quadrant by these amounts ($\Delta r$ = 0.01 mag, $\Delta i$ = 0.23 mag, $\Delta z$ = 0.13 mag). Since the $i$ and $z$ magnitudes (which were observed interleaved together) are most significantly discrepant, it is likely that some of the cirrus observed earlier in the night was in front of our field during these observations. We used these offsets with overlapping stars to correct the magnitudes in this quadrant.

\subsection{UKIDSS Data}
\label{ukidss}
We obtained UKIDSS Data for our two regions from the WFCAM Science Archive\footnote{\url{http://surveys.roe.ac.uk/wsa/index.html}}. B5 is covered in the Galactic Clusters Survey (GCS) while most of the West-End is covered in the Galactic Plane Survey \citep[GPS;][]{Lucas:2008}. We used the following quality cuts in the SQL query as recommended by \citet{Lucas:2008} for high-reliability photometry:
\begin{verbatim}
jAperMag3 > -10 and hAperMag3 > -10 and 
k_1AperMag3 > -10 and jAperMag3Err < 0.03 
and hAperMag3Err < 0.03 and k_1AperMag3Err < 0.03 
and jEll < 0.2 and hEll < 0.2 and k_1Ell < 0.2
and jpperrbits < 256 and hpperrbits < 256 and 
k_1pperrbits < 256 and pstar > 0.99 and 
sqrt(hXi*hXi + hEta*hEta) < 0.3 and 
sqrt(k_1Xi*k_1Xi + k_1Eta*k_1Eta) < 0.3
and mergedClass !=0 and 
(PriOrSec=0 or PriOrSec=framesetID)
\end{verbatim}
and cross-matched the Megacam data with the UKIDSS data for stars within 0.6\arcsec. The UKIDSS survey data is approximately on the 2MASS colour system, although the filters and system responses are slightly different. We discuss the influence this may have on our results in Section~\ref{biases}. The UKIDSS data is preferable to the 2MASS data because the uncertainties on 2MASS colours for our stellar sample are many times larger than the uncertainties on our Megacam data and so 2MASS errors would dominate the errors in our analysis.

\section{Fitting Procedures}
\label{section:fitting}
\subsection{Least-Squares Fitting Approach}
\label{Least-Squares}
The traditional least-squares method for determining correlation between \av\ and \rv\ involves first deriving best-fit estimates for each star's \av\ and \rv\ and then testing these best-fit values for correlation with another least-squares procedure.

We formalise the details of our fitting procedure as follows. Our input data are five observed colours $O_{n}$ derived from 6 photometric bands  and their associated errors $\sigma_{O_{n}}$. Our set of colours ($O_{n}$) is ($r-i$,$i-z$,$z-J$,$J-H$, $H-K$). Our model relies on an empirically-derived, fifth-order polynomial parameterization of the intrinsic colours of the stellar locus, $C_{n}$:
\begin{equation}
C_{n} = \sum_{k=0}^5 D_{k,n} x^k
\label{stellarlocus}
\end{equation}
where the parameters, $D_{k,n}$ defining each locus come from \citet{Covey:2007}, and are based on an examination of the SDSS and 2MASS colours of 600,000 point sources at low extinction values. The equation in \citet{Covey:2007} uses $g-i$ colour in Eqn.~\ref{stellarlocus} as a proxy for stellar type. We use the variable $x \equiv g-i$ instead, to emphasize that we do not use $g$-band data in this study. Eqn.~\ref{stellarlocus} simply provides a means to convert a single parameter ($x$, the intrinsic stellar type) into our observed colours (the set $C_{n}$).

This stellar locus has finite (observed) width, which is due to both intrinsic and measurement dispersion. This width is different for each colour, $n$, and varies along the sequence (i.e. is a function of intrinsic stellar type, $x$). We average this dispersion along $x$ to produce a single number characterising the width of the observed stellar locus, $\sigma_{C_{n}}$, and we assume that the intrinsic width of this distribution is one-half the reported width, to crudely account for the difficult-to-assess influence of measurement error.

\begin{figure}
\includegraphics[width=8.5cm]{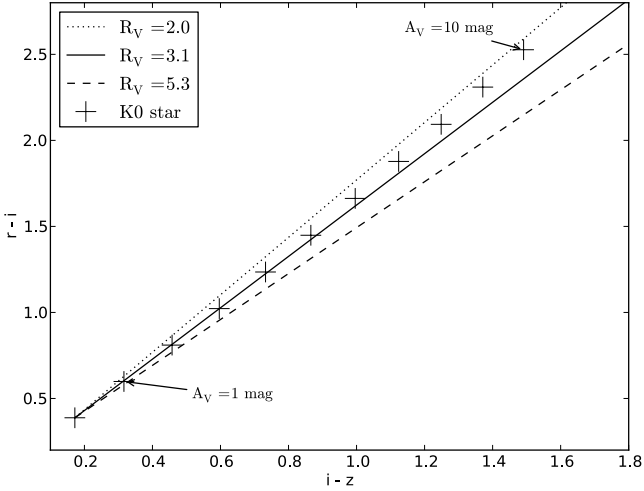}
\caption{Red-optical ($r-i$ versus $i-z$) colours of a \citet{Castelli:2004} model K0V star produced in {\sc synphot} by convolving the stellar spectrum with the filter response curves for the MMT. The stellar spectrum is reddened with a CCM \rv\ = 3.1 reddening law in steps of \av\ = 1 magnitude (plus symbols). The change in the effective wavelength of the reddened spectrum causes the colours to trace out a curved line in colour-colour space, rather than the straight vectors assumed in our model (straight lines). This effect makes it look as if a star behind more extinction is being reddened by an extinction law with a smaller \rv.}
\label{CurvedExtinction}
\end{figure}

Our model also includes extinctions, $E_{n}$, where
\begin{equation}
E_{n} = A_{\lambda_1} - A_{\lambda_2}
\end{equation}
\begin{equation}
A_{\lambda_1} = A_V (a_{\lambda_1} + b_{\lambda_1} R_V^{-1})
\end{equation}
and $\lambda_1$ represents the effective central wavelength for a particular band. The coefficients $a_{\lambda_1}$ and $b_{\lambda_1}$ are taken to be constant for each band, although formally, this is an integration across the band which is contingent on the stellar spectrum (parameterized in our model by $x$), the filter response, and the atmospheric opacity at the time ($t$) of each observation :
\begin{equation}
A_{\lambda_1} = \int_{\Delta{\lambda_1}} \textrm{Filter}(\lambda) \times \textrm{Atmosphere}(\lambda,t) \times \textrm{SED}(\lambda,x) d\lambda.
\end{equation}
The extinction in a given waveband is therefore not a single number, but a function of the intrinsic stellar SED and the amount of extinction that starlight has already passed through. It is more correct to think of reddening curves rather than vectors \citep[][uses the term `reddening track']{Stead:2009}. Figure~\ref{CurvedExtinction} shows this effect by plotting the $r-i$ versus $i-z$ colours of a K0~V stellar spectrum with increasing amounts of extinction assuming a CCM law with \rv\ = 3.1 (produced with {\sc synphot}\footnote{\url{www.stsci.edu/resources/software_hardware/stsdas/synphot}}). The extinction vectors plotted are appropriate for an un-reddened stellar spectrum, but as the spectrum undergoes increasing amounts of extinction the effective wavelength changes. As \av\ increases, the resulting stellar colours describe a curve, rather than a straight vector. 

Figure~\ref{CurvedExtinction} shows the worst case in our study, since it considers the $r-i$ and $i-z$ colours which are most sensitive to reddening and includes up to 10 magnitudes of visual extinction, which is the maximum value we infer for any stars in this study. Most stars are behind only a few magnitudes of \av\ or less, and the longer-wavelength colours are less influenced. Including this effect would increase the complexity of the model significantly since we would have to iteratively compute the reddening. Therefore we do not include it for this study (but see Section~\ref{biases} for a discussion of systematic biases and other possible errors), where the column of dust is relatively small ($<$ 10 magnitudes of extinction in \av), but it should not be ignored when considering the extinction law in denser regions. 

We evaluate $a_{\lambda}$ and $b_{\lambda}$ from the CCM model. Defining $\xi \equiv \lambda^{-1}  \mu m^{-1}$, for $\xi > 1.1$ (i.e. for $r$ and $i$)
 \begin{equation}
 y \equiv \xi - 1.82,
 \end{equation}
 where the numerical factor of 1.82 is simply for convenience,
\begin{equation}
a_{\lambda} = 1.0 + 0.177 y - 0.504 y^2 - 0.024 y^3 + 0.721 y^4 + 0.020 y^5 - 0.775 y^6 + 0.330 y^7
\end{equation}
\begin{equation}
b_{\lambda} = 1.413 y + 2.283 y^2 + 1.072 y^3 - 5.384 y^4 - 0.623 y^5 + 5.303 y^6 - 2.090 y^7
\end{equation}

while for $\xi < 1.1$ (i.e. for $z$, $J$, $H$, $K$)
\begin{equation}
y \equiv \xi^{1.61}
\end{equation}
\begin{equation}
a_{\lambda}  = 0.574 y
\end{equation}
\begin{equation}
b_{\lambda} = -0.527 y.
\end{equation}
These coefficients are summarised in Table~\ref{CCM Coeffs}.

\begin{table}
\begin{tabular}{lccc}\hline
Filter & $\lambda_{eff}$ (\AA) & a$_\lambda$ & b$_\lambda$ \\\hline
$r$ &  6374 & 0.930 & -0.240 \\
$i$ &  7797 & 0.799 & -0.533 \\
$z$ & 9052 & 0.675 & -0.619 \\
$J$ & 12554 & 0.414 & -0.380 \\
$H$ & 16358 & 0.260 & -0.239 \\
$K$ & 22002 & 0.162 & -0.149 \\\hline
\end{tabular}
\caption{\label{CCM Coeffs}CCM Coefficients}
\end{table}

We then attempt to minimize (using {\sc mpfitfun}\citep{Markwardt:2009}, aka {\sc minpack-1}) for each star the quantity
\begin{equation}
M = \sum_{n} \left[\frac{O_{n} - C_{n}(x) + E_{n} (A_V,R_V)}{\sqrt{\sigma^2_{O_{n}}+\sigma^2_{C_{n}}}}\right]^2
\label{lsq-eq}
\end{equation}
with the constraints that \av\ $>$ -0.5 mags, 2.1 $<$ \rv\ $<$ 5.5, and 0 $< x <$4.5 (recall that $x \equiv g-i$ and parameterizes the spectral type of the star). We allow \av\ to go slightly negative because the intrinsic $C_{n}$ are derived from fitting real stars with some small amount of \av. 

The problem with this approach is that we often run into limits on \rv. Relaxing the \rv\ limits does not help; there is no local minimum in $\chi^{2}$-space along this dimension. In particular, in the case of low \av, we have very little handle on \rv, and not a great deal on \av\ (the intrinsic spectral type, $x$, is normally fit robustly in such a case). The large number of points which hit our limits on \rv\ make it hard to test the hypothesis that Cov($A_V$,$R_V$) $\neq 0$ for confidence. Another statistical drawback of this approach is that applying standard statistical estimators to a sample of estimates in the presence of error results in biased estimates of the population variance and correlation \citep[e.g.][]{bkelly07,loredohendry10}.  That is, when each individual star's $(A_V, R_V)$ parameters are fitted separately, the resulting ensemble of estimates will be wider than the intrinsic variance, and the apparent correlation of the ensemble will be diminished relative to the true intrinsic correlation.

\begin{figure}
  \begin{center}
    \begin{tabular}{cc}
      \resizebox{3.8cm}{!}{\includegraphics[angle=0]{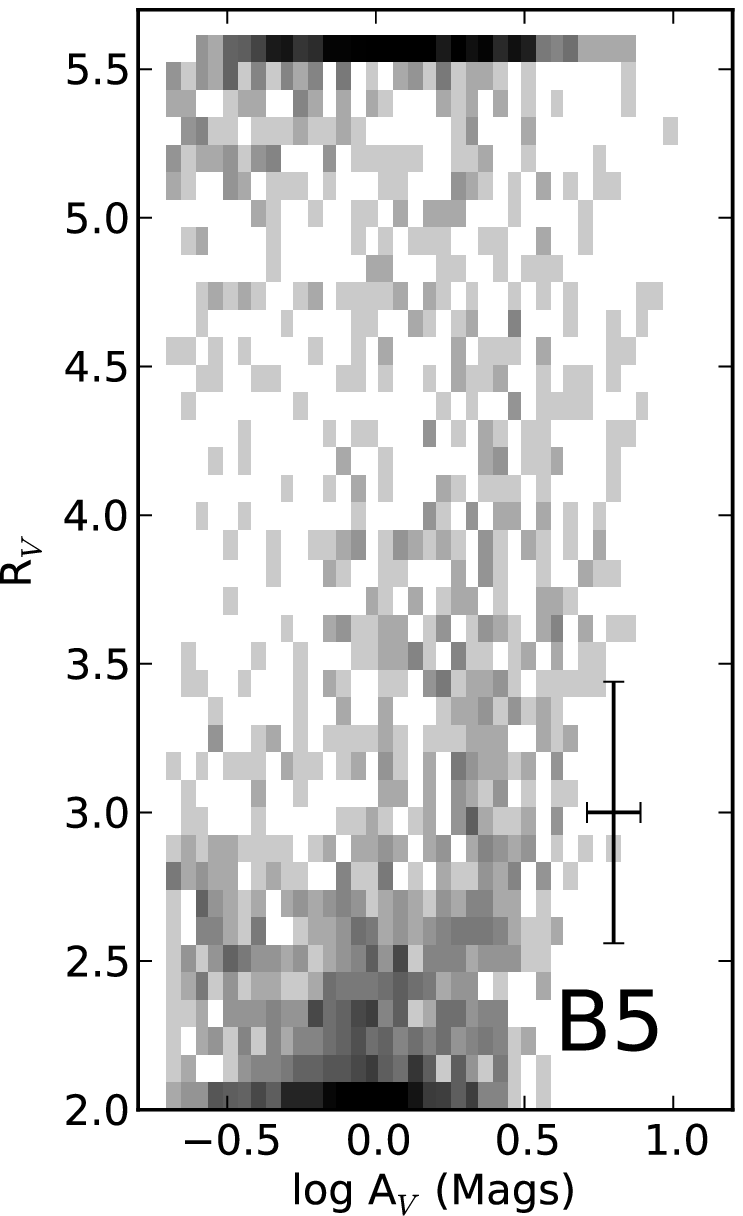}}&
     \resizebox{3.8cm}{!}{\includegraphics[angle=0]{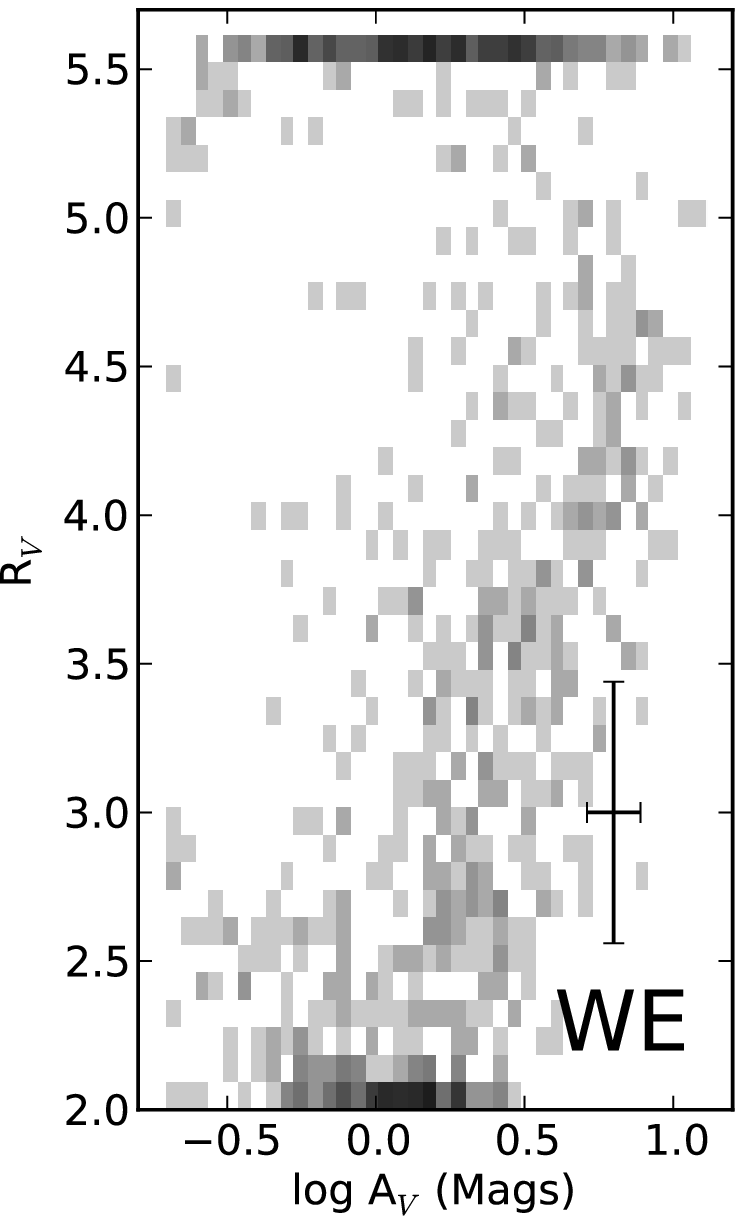}}
    \end{tabular}
\caption{Least-squares fits for \av\ and \rv\ for the stars in our data samples for B5 (left panel) and the West-End (right panel). The shades of grey show the density of points at a particular location in the diagram. A representative error bar is shown, which displays the median uncertainty on \av\ and \rv\ for a single star (rather than for one of the bins displayed in this figure). Stars for which no local $\chi^{2}$-minimum exists between \rv\ = 2 and \rv\ = 5.5 ``pin'' at these extreme values. These points are therefore unreliable.}
\label{LSQ}
  \end{center}
\end{figure}

We plot these results for both B5 and the West-End in Figure~\ref{LSQ}. Despite the large number of points which hit the limit, there is a suggestion of a correlation in the remaining points for values of \av\ above 1 magnitude. This correlation is in the direction expected---larger values of \rv\ at higher column density. 

\subsection{A Hierarchical Bayesian Approach}
\label{section:bayesian}
We developed a hierarchical Bayesian model to address some of the problems present in the least-squares fitting approach. This model allows us to constrain the properties of the dust towards, and the intrinsic colours of, \emph{individual} stars by simultaneously modelling the \emph{populations} of dust properties and stellar colours. These parameters of the full population are called hyper-parameters, and describe the shape of the distributions of $A_V$ and $R_V$. An advantage to this approach is that we can include the correlation of $A_V$ and $R_V$ explicitly in our model as a hyper-parameter, and then examine its marginal posterior probability distribution conditional on our full data-set in order to make probabilistic inferences about the dust population. Furthermore, by taking a fully Bayesian approach, we can coherently compute the joint uncertainties in the estimates of both the hyper-parameters governing the population and the dust and colour parameters of individual stars.   By modelling the intrinsic population distribution directly, the hierarchical approach coherently accounts for the relevant estimation errors when inferring the intrinsic population covariances, thus correcting the biases that plague standard estimators in the presence of error.  The mathematical details of the statistical model are given in Appendix \ref{appendix:model}.  All other aspects of the dust model are as described in \S~\ref{Least-Squares}.  The statistical and numerical methods used here are similar in principle to those introduced by \citet{Mandel:2009,Mandel:2011} for estimating dust extinction to Type Ia supernovae using optical and near-infrared photometry.

The distribution of column densities ($A_V$) within a molecular cloud is often observed to be log-normal \citep[e.g.][]{Goodman:2009a}, though this is not always true. For example, \citet{Lombardi:2008} find a log-normal distribution for Ophiuchus and Lupus, but \citet{Lombardi:2006} see a more complex density distribution for the Pipe nebula, possibly due to multiple clouds along the line of sight. Many simulations of turbulent molecular clouds \citep[e.g.][]{Vazquez-Semadeni:1994,Ostriker:2001} predict a log-normal volume density distribution. This is equivalent to a log-normal column density distribution if the column density along each line of sight through the cloud is dominated by a single feature. We therefore take a log-normal distribution as a prior on $A_V$:
\begin{equation}
\log(A_V) \sim N(\mu_A, \sigma^2_A).
\end{equation}
The distribution of $R_V$ values within a molecular cloud is unknown. However, in the range of column densities probed in this study we expect $R_V$ to occupy a range of values peaked near the value for the diffuse ISM ($R_V$ = 3.1) or slightly larger. A Gaussian in the inverse of $R_V$ (\rr\ $\equiv R^{-1}_V$) is a particularly mathematically convenient choice:
\begin{equation}
\mrr \sim N(\mu_r, \sigma^2_r).
\end{equation}
We assume a flat (uninformative) prior for the intrinsic spectral type, $x$, between 0.2 and 4.2, as our experience suggests that this is a parameter well-constrained by the data. See \S~\ref{biases} for a further discussion of this choice. Our probability model for the stellar colours is that each measured colour comes from a normal distribution, with the mean colour vector (conditional on the intrinsic spectral type, $x$) given by a 5th order polynomial (Eq.~\ref{stellarlocus}), and the standard deviations in each colour given by a 5th order polynomial derived from the uncertainties quoted in \citet{Covey:2007}.

We further assume the intrinsic spectral type, $x$, is not \emph{a priori} correlated with the other parameters in the population, although a weak correlation is possible with $A_V$ because only bright blue stars are visible through high column density material. We aim to test the hypothesis that \av\ and \rr\ have a non-zero population correlation, which we express as
\begin{equation}\label{eqn:binormal}
\left(\begin{array}{c} \mrr\ \\ \log(A_V) \end{array}\right)  \sim N\left[\left(\begin{array}{c} \mu_r \\ \mu_A \end{array} \right), \left(\begin{array}{cc} \sigma^2_r & \rho\ \sigma_A \sigma_r \\ \rho\  \sigma_A \sigma_r & \sigma^2_A \end{array} \right)\right].
\end{equation}
In fact our model could be extended to describe a general polynomial dependence between \rr\ and \av. However, we restrict ourselves to a linear dependence, in which case the joint population distribution of $\log(A_V)$  and \rr\ is sufficiently described by a linear correlation parameter $\rho$. The exact form of this linear relationship is
\begin{equation}
\log{A_V} = \alpha_{0}+\alpha_{1} (r_{V} - 0.32)/0.04 + \epsilon
\label{alphas}
\end{equation}
where $\alpha_{0}$ and $\alpha_{1}$ are the intercept at $R_V = 3.1$ and the slope of the linear relationship, respectively.  The numerical constants are simply for convenience.  The residual variance $\text{Var}[\epsilon] = \sigma^2$, and $\bm{\alpha}$ are related to the mean and covariance hyper-parameters of Eq. \ref{eqn:binormal}, via Eqs. \ref{eqn:conversions}.

\begin{figure}
\centering
\includegraphics[width=8cm]{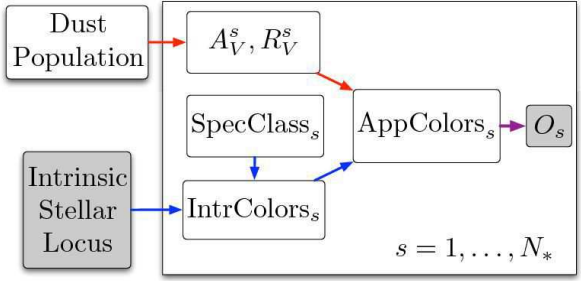}
\caption{\label{fig:dag} The global posterior density of the unknowns given the full data set of apparent stellar colours is represented by a directed acyclic graph. Unknown parameters are represented by open nodes. Observed data (measured colours $\bm{O}$) and knowns are represented by shaded nodes. The directed arrows between the nodes indicate relations of conditional probability. The hierarchical model coherently incorporates randomness and uncertainties due to measurement error (purple arrows), spectral class ($x = g-i$) and the intrinsic stellar locus ($\bm{\mu}_C(x), \bm{\Sigma}_C(x)$) (blue arrows), and dust extinction and reddening $(A_V, R_V)$, and its population ($\bm{\alpha}, \sigma^2, \mu_r, \sigma_r^2$) (red arrows) into inferences about individual stars and the population. The probabilistic graphical model describes a conceptual mechanism for generating the observed data.}
\end{figure}

Again representing the intrinsic stellar colour as $x \equiv g-i$, and denoting the parameters and data for each of $N$ stars with the label $s$, the suite of hyper-parameters as $\mathbf{H}$ = [$\mu_r, \sigma^2_{r}, \bm{\alpha}, \sigma^2$], and the vector of observed colours for each star as $\mathbf{O^s}$, we seek to compute the global posterior distribution

\begin{equation}
\begin{split}
P(& \{A_V^s, r_v^s, x_s\}; \mathbf{H} |\, \{ \mathbf{O^s} \}) \propto \\
& \prod_{s=1}^{N} \bigg[ P(\mathbf{O^s} | x_s, A_V^s, r_v^s) \times P(x^s,A^s_V,r^s_v | \mathbf{H}) \bigg] \times P(\mathbf{H}),
\end{split}
\end{equation}
where we place uniform (non-informative) priors, $P(\mathbf{H})$, on the hyper-parameters.  The derivation of this expression is described in detail in Appendix \ref{appendix:model}.  The probabilistic structure of the statistical model is shown as a directed acyclic graph (DAG) in Fig. \ref{fig:dag}.  In this graphical model, the unknown parameters are depicted with white boxes.  The knowns and observed data that the inferences are conditioned upon are described by grey boxes.  The relationships of conditional probability between all the variables (specified in detail in \S\ref{appendix:model}) are shown as directed arrows. The graph can be thought of as a \emph{generative} model, or conceptual mechanism for producing the observed data.   Given the intrinsic stellar locus, for a given spectral class ($x_s$), a set of intrinsic stellar colours is generated.  From the dust population, random values of the dust parameters ($A_V, R_V$) are drawn.  The intrinsic colours and dust effects combine to produce the apparent colours, which are then sampled with measurement error to generate the observed colours $\bm{O}_s$.
Further information on DAGs and examples of their application in astronomy can be found in \citet{Mandel:2009,Mandel:2011,bishop}.

Because we are considering several thousand stars at once, there is a large number of parameters and hyper-parameters to be estimated jointly ($3N+5$). Details of an efficient algorithm for solving this problem are presented in Appendix \ref{appendix:mcmc}.  
A Markov Chain Monte Carlo algorithm with generalised Gibbs sampling was used to efficiently explore the probability distribution of the model parameters. This algorithm was designed to avoid getting stuck at spurious local maxima with extreme \rv\ (the problem plaguing the least-squares fitting approach), and to negotiate multiple possible solutions in the global posterior distribution. At each step we draw a new value of a given parameter by sampling from the probability distribution of this parameter, conditioned on the current value of all the other parameters at this step. In each full cycle we sample the intrinsic colour ($x$), \av\ and \rr\ for each star, and then sample the hyper-parameters which describe the whole population. After a full cycle, the current value of all parameters is recorded as the position of the Markov chain. After a sufficient number of iterations the Gibbs sampler converges to a stationary solution and maps out the full posterior probability distribution of parameters and hyper-parameters conditioned on the full data set of observed colours. 

We used least square fitting (Eq.~\ref{lsq-eq}) to estimate initial values for each parameter for each stars. To each value we added some random noise to generate different starting positions for each individual Markov chain. We remove the first 10\% of each chain as ``burn-in'', during which time the chain is still seeking the equilibrium distribution of the parameters. We sample 2500 times with 4 parallel, independent chains, but record only every 10th sample to reduce the autocorrelation within each chain. We measure the convergence of the model using the Gelman-Rubin \citep{Gelman:1992} statistic. This compares the variance between chains and within chains to estimate the mixing of the chains and diagnose their convergence to the equilibrium posterior distribution. We consider a chain to have converged if the maximum Gelman-Rubin ratio is less than 1.1. 

This general approach in which we use population information to infer properties of individual stars in a Bayesian context is similar to the work reported in \citet{Bailer-Jones:2011}. Unlike that work, our model does not include parallax information. We make no attempt to solve for distance, considering only colours rather than magnitudes. In addition, our model does not consider prior information about the expected density of stars of different spectral types; we consider all spectral types ($g$-$i$) to be equally probable. Our study incorporates a more detailed model for the dust extinction as that is the focus of this work.

In some respects this work is similar to \citet{Kelly:2012}. \citet{Kelly:2012} use a hierarchical Bayesian approach to infer changes in dust grain properties as a function of dust temperature and column density in the Bok Globule CB244. \citet{Kelly:2012} examined the dust emissivity spectral index ($\beta$) in the far infrared and millimetre and found, in contrast to many prior studies, no anti-correlation between $\beta$ and temperature, but a strong correlation between $\beta$ and column density. In that work, each independent pixel in an emission map was fit individually, and the overall population modelled with hyper-parameters. In our work, each individual star is fit individually, and the overall population is modelled with hyper-parameters. Future work could combine the information about dust emission in \citet{Kelly:2012} with the information in this study about dust extinction to more tightly constrain models of dust growth.

\subsection{Testing the Gibbs Sampler}
\label{testing}
To test our hierarchical model for sensitivity to factors which could lead to spurious correlations, we constructed fake data sets and assessed our ability to recover input parameters. Un-reddened stars were drawn from the empirical distribution of \citet{Covey:2007} using the observed distribution of intrinsic $g-i$ colours as the underlying probability distribution. This basic $g-i$ colour was used to generate the other five colours ($r-i$, $i-z$, $z-J$, $J-H$, and $H-K$) by drawing from Gaussian distributions with means and standard deviations determined by the stellar colour loci and standard deviations tabulated in \citet{Covey:2007}. Values of  $\log(A_V)$ and \rr\ were drawn from specified distributions with known hyper-parameters and used to redden stellar colours according to our CCM parameters. Gaussian errors of a specified magnitude were then added to simulate the observed colours.

Our primary goal is to constrain $\rho$, the correlation between $\log(A_V)$ and \rr. To test this, we generated test data sets, each with 2000 stars. Values of $\log(A_V)$ and \rr\ were drawn from joint normal populations in order to produce data with a particular correlation. Because of finite sampling, these simulations did not produce a population with exactly the specified correlation, so we measured the sample correlation in these parameters after generation and use that value for comparison. The marginal posterior distributions on most hyper-parameters were approximately normal, with mean values close to (well within the 95\% interval) the input values. As would be expected, the posterior distributions for the parameters of individual stars covered the input values---we had to recover the input values of individual stars reasonably well in order to get the population parameters correct. Many stars had very broad posterior distributions on \rv, as we expected.

As a basic test of our sensitivity to the mathematical form of our prior distributions, we conducted another series of tests in which $\log(A_V)$ and  \rr\ were drawn from a bivariate uniform distribution with a linear correlation over a sensible range ($A_V$ from 0.1 to 10 mag, \rv\ from 2.0 to 5.5). Again, $\rho$ was recovered accurately. 

\subsection{Potential Biases}
\label{biases}

We know that our model contains some important simplifications which may bias our result. In addition, our catalogue of stellar colours has been produced using some simplifying assumptions that could, since we are considering small colour shifts, bias our result. We discuss what we believe are the most important assumptions and sources of bias here. 

For each filter ($r$, $i$ ,$z$ ,$J$ ,$H$ ,$K$) we assume a single effective wavelength for all stars when calculating the parameters in Table~\ref{CCM Coeffs}, but this assumption is not precisely true. However, the spread in effective wavelengths for the stellar population we study (which is dominated by late-type stars) is small. The spread in effective wavelength is largest in $z$, where $\Delta \lambda_{eff} = 15$\AA\ between an F0 and K0 star. 

A more significant shift in the effective wavelength is produced by reddening itself. As seen in Figure~\ref{CurvedExtinction}, when extinction reaches many magnitudes of visual extinction, the extinction tracks are actually curves rather than straight vectors (as we assume). This introduces a bias in our work, but crucially it is the opposite direction of the effect we infer in \S~\ref{sec:results}. That is, if we were to calculate \rv\ from Figure~\ref{CurvedExtinction} for the \av\ = 10 point, we would infer a value of \rv\ $< 3.1$, closer to \rv\ = 2. Since all stellar spectra behave in qualitatively the same way under extinction (effective wavelengths become longer in each band, but more quickly in blue bands), this effect would tend to bias us toward inferring that \rv\ decreases with \av, the opposite of what we actually infer. The effect is small for the low extinction we study here, and thus probably has only a small impact biasing our results toward finding less of a correlation between \rv\ and \av.

\citet{Naoi:2007} point out the dangers of photometric transformations in studies which attempt to study the extinction law (they also find a change in the extinction law as a function of increasing optical depth). A similar conclusion is reached by \citet{Gosling:2009} in their study of near-infrared extinction toward the nuclear bulge and \citet{Kenyon:1998} in their study of Taurus---their results are only properly valid in their own photometric system. We have attempted to transform our Megacam data onto the SDSS system and the UKIDSS data is calibrated onto the 2MASS system. These transformations are necessary, since we use an observed stellar colour locus in SDSS and 2MASS colours. Despite the large number of stars we are able to use for these transformations, this transformation probably remains our largest source of systematic uncertainty. 

Our prior on intrinsic spectral type (parameterized by $x \equiv g-i$) is relatively uninformative. That is, we use a uniform prior between a range of intrinsic $g-i$ colours. The true distribution of intrinsic spectral types is probably not uniform, but we do not have a good estimate for this distribution \emph{a priori}. In \S~\ref{testing} we use the observed $g-i$ colour distribution from \citet{Covey:2007} when testing the Gibbs sampler, but this distribution is for the portion of the Galaxy sampled by the low-extinction SDSS footprint; it is not necessarily appropriate for the particular portion of the Galactic stellar population sampled in this study. Even with our uninformative prior, the median posterior uncertainty on the intrinsic $g-i$ colour is only 0.24, significantly smaller than the width of allowed colours in our prior ($0.2 < g-i < 4.2$). We choose to leave the prior relatively uninformative, rather than use a potentially incorrect prior. 

We also do not assume that there is an \emph{a priori} correlation between intrinsic stellar type and \av. This correlation may be is present at large column densities where only the brightest background stars and foreground stars are detected. In extinction studies, methods such as {\sc NICEST} improve on the NIR colour excess method ({\sc NICE}) to account for the bias that this correlation produces in extinction maps \citep{Lombardi:2009}. As noted in \citet{Lombardi:2009}, the bias in extinction maps only becomes significant at column densities where $A_{V} >  10$ magnitudes, which is the maximum extinction we probe. For this reason, we do not attempt to model this correlation. 

Additional possible systematic effects include calibration problems with our Megacam data as we saw in the fourth quadrant of B5 and correlations in the underlying population not represented in our model. Our study fits slope of the extinction law under the parameterization of the CCM model. If the extinction in these portions of Perseus is not well described by the CCM model then our fits to the CCM model may not be particularly enlightening. A number of objects in our survey may not have intrinsic colours lying on the \citet{Covey:2007} tracks. These include unresolved background galaxies or quasars, embedded young stars and brown dwarfs. These objects probably constitute a small fractional contamination ($\sim 10\%$) at the depths studied in this work \citep[see the estimates in][]{Foster:2008}.

\section{Results}
\label{sec:results}
We ran our two regions separately, but the results are similar. In each case, a few objects failed to converge, producing large maximum values of the Gelman-Rubin statistic. Other than that, the chains behaved well, with Gelman-Rubin less than 1.1 for 99.9\% of the objects in both regions. We identified the objects which failed to converge, removed them from the catalogue, and re-ran the model. Doing this  verified that the inferences on the hyper-parameters were unaffected by these objects.

\begin{figure}
  \begin{center}
    \begin{tabular}{cc}
      \resizebox{3.8cm}{!}{\includegraphics[angle=0]{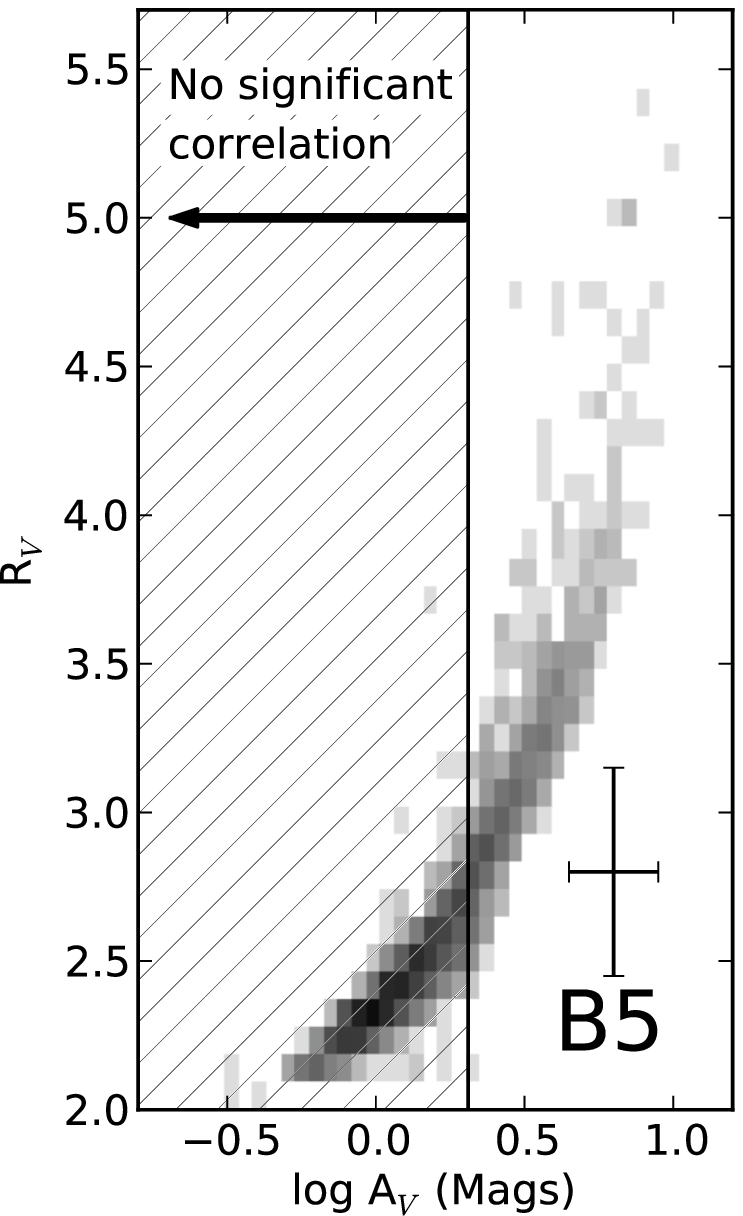}}&
     \resizebox{3.8cm}{!}{\includegraphics[angle=0]{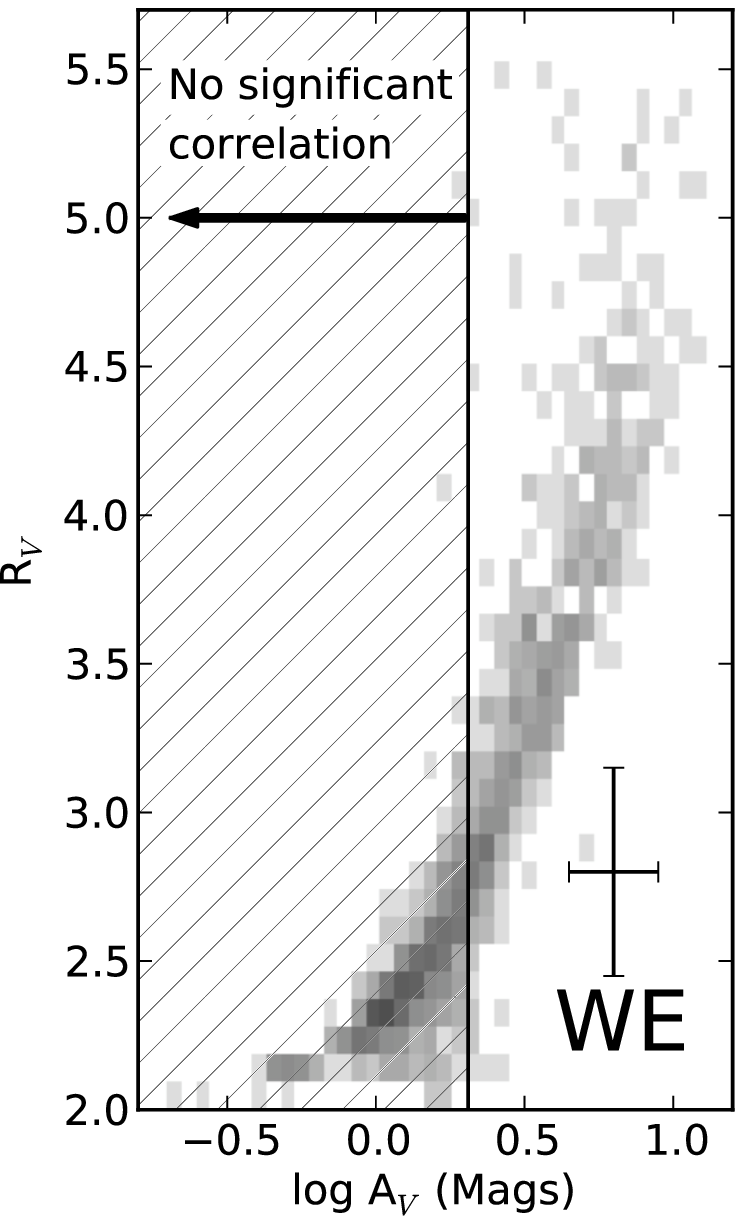}}
    \end{tabular}
\caption{The mean and standard deviations of the marginal posterior distributions on \av\ and \rv\ for all the stars in our data samples for B5 (left panel) and the West-End (right panel) from our hierarchical Bayesian model. The shade of grey shows the density of points at a position within the diagram. A representative error bar is shown, which displays the median width of the posterior probability distribution for \av\ and \rv\ for a single star (rather than for one of the bins displayed in this figure). Note that the widths of the posterior probability distributions for \av\ and \rv\ vary greatly among individual stars. When we include only sources with \av\ $<$ 2 magnitudes the correlation between \av\ and \rv\ is no longer distinguishable from zero. For these points we are unable to infer any underlying correlation and the tight relationship seen in this diagram is due to the strong correlation inferred for stars with \av\ $>$ 2 magnitudes.}
\label{AvRv}
  \end{center}
\end{figure}

\begin{figure*}
  \begin{center}
    \begin{tabular}{cc}
      \resizebox{8.5cm}{!}{\includegraphics[angle=0]{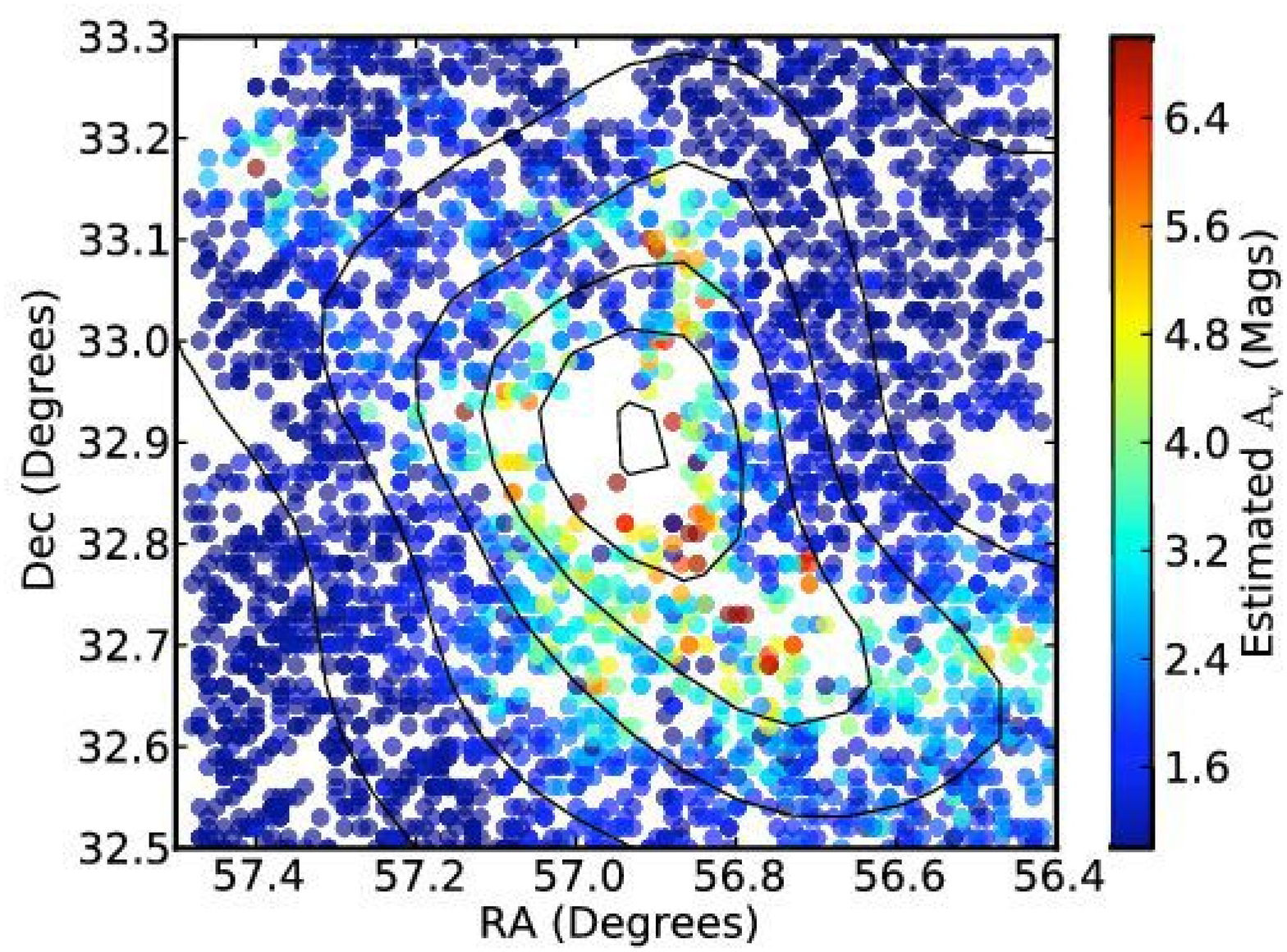}}&
     \resizebox{8.5cm}{!}{\includegraphics[angle=0]{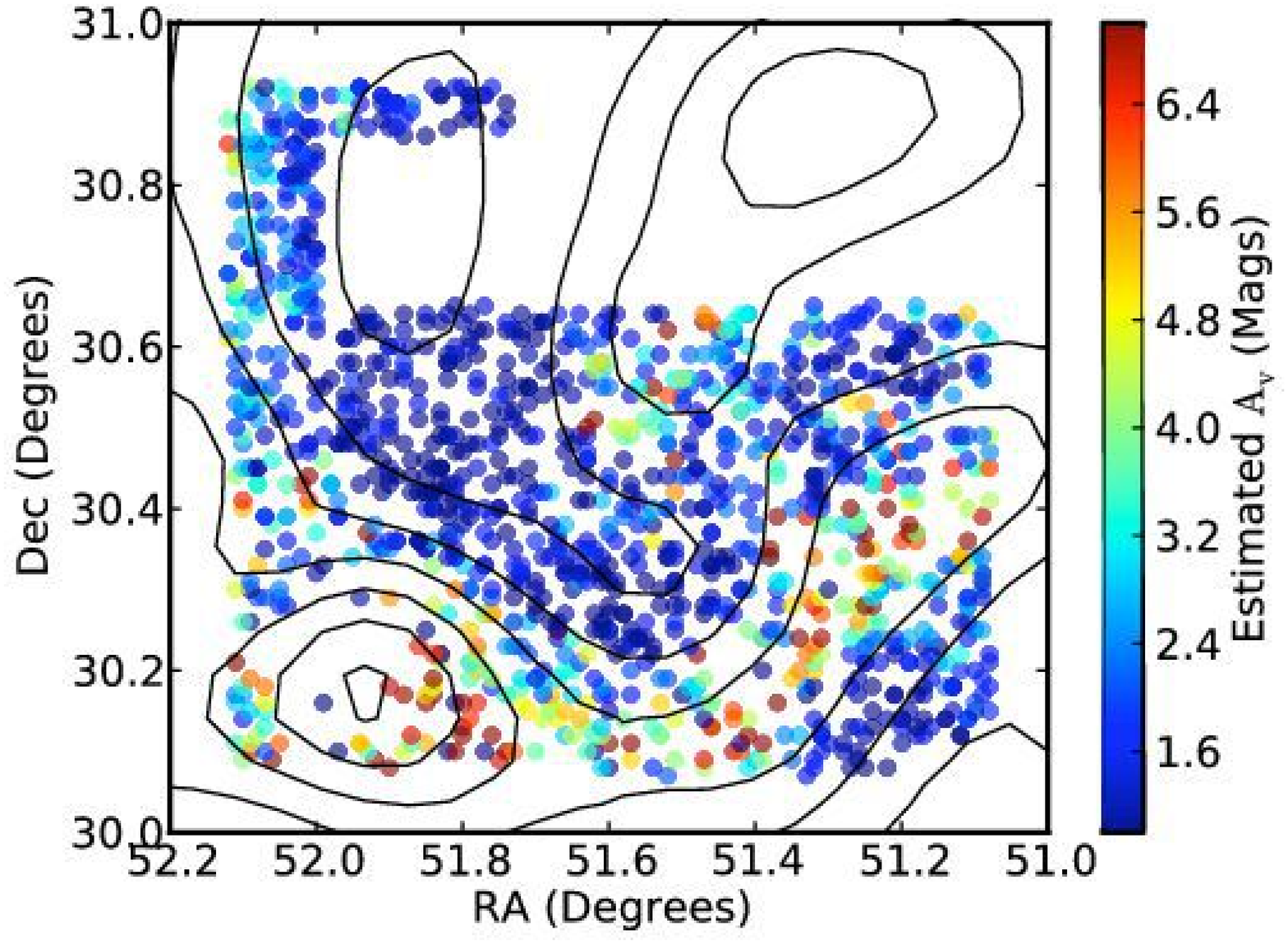}}
    \end{tabular}
\caption{Spatial distribution of inferred \av\ in B5 (left panel) and the West-End (right panel) using our hierarchical model with extinction coded in colour. Overlain are \av\ contours from the COMPLETE NICER map \citep{Ridge:2006} with contour levels at 1.6, 2.4, 3.2, 4.0, 4.8, 5.6 and 6.4 magnitudes of \av. The spatial distribution of stars with high extinction from our Bayesian analysis conforms to known cloud structures. The lower density of points in West-End is principally the UKIDSS/GPS data available for this region is less deep than the UKIDSS/GCS data for B5. Gaps at the edges can be due to non-uniform coverage either in our Megacam observations (e.g. B5) or in the UKIDSS data (e.g. West-End). Gaps in the stellar distribution in regions of high extinction are areas where the extinction is too high for background stars to be seen in our Megacam observations.}
\label{Positions}
  \end{center}
\end{figure*}

The inferences on \rv\ and \av\ are shown in Figure~\ref{AvRv}. These plots show the same basic behaviour in both samples, with \rv\ rising with \av. The points at low \av\ are poorly constrained by the data from the individual stars, and the posterior estimates of \rv\ and \av\ in this regime are strongly informed by the population hyper-parameters. 

To assess the reliability of inferences at low \av\ we performed a series of tests in which we progressively removed stars with large inferred values of \av. We used our initial catalogue of inferred values for \av\ to remove the most highly extinguished stars and reran the analysis with these subsets of low-extinction stars. The posterior probability distribution on $\rho$ becomes broader until, when considering only stars with \av\ $<$ 2 magnitudes, the inference on $\rho$ no longer is inconsistent with zero. That is, if we consider only stars with \av\ $<$ 2 magnitudes, we are not able to claim that a significant correlation exists between \rv\ and \av. The inference on \rv\ is very weak for these stars because there is not a significant amount of reddening. Ultimately, therefore, the \rv\ and \av\ inferences for these stars is strongly determined by the (linear) relationship we assume between \rr\ and $\log(A_V)$ and the hyper-parameters which describe this relationship. Without the stars at high extinction we would have very little information about the relationship between \rv\ and \av. One should therefore not pay too much attention to the relationship below \av\ of 2 magnitudes. This result gives us some confidence that subtle errors in catalogue colours (due to errors in calibration or photometric transformations as discussed in \S~\ref{biases}) are not producing a spurious relationship; the lack of strong correlation at low \av\ makes it more likely that the strong correlation we infer at high \av\ is genuine.

An alternative approach would be to expand our model to allow for a non-linear population relation between $\log A_V$ and $r_V$, with which the strength of the correlation could change as a function of $\log A_V$.   This would require building a more advance statistical model and significant changes to the sampling algorithm, which we leave for future work.

\begin{table*}
\begin{tabular}{llccccccc}\hline
& \# Stars & $\mu_r$ & $\sigma_r$ & $\mu_A$ & $\sigma_A$ & $\alpha_{0}$ & $\alpha_{1}$ &  $\rho$ \\\hline
B5 & & & & & & & & \\
&3144 & 0.391$\pm$0.002 & 0.062$\pm$0.001 & 0.34$\pm$0.01 & 0.31$\pm$0.01 & 0.95$\pm$0.02 & -0.338$\pm$0.008 & -0.86$\pm$0.01 \\
West-End & & & & & & & & \\
& 1278 & 0.363$\pm$0.003 & 0.082$\pm$0.002 & 0.63$\pm$0.02 & 0.38$\pm$0.01 & 0.94$\pm$0.02 & -0.287$\pm$0.008 & -0.84$\pm$0.01 \\
\hline
\end{tabular}
\caption{\label{HyperParameters} Median Marginal Posteriors of Hyper-parameters. $\mu_r$ and $\sigma_r$ are the hyper-parameters describing the normal distribution of \rr, which is $R^{-1}_V$; $\mu_A$ and $\sigma_A$ describe the normal distribution of $\log(A_V)$. $\alpha_0$ and $\alpha_1$ describe the linear relationship between \rr\ and $\log(A_V)$ as defined in Eqn.~\ref{alphas}. $\rho$ is the correlation between \rr\ and $\log(A_V)$. 
}
\end{table*}

We compute the posterior distributions of individual hyper-parameters marginalised over all the other parameters. These posterior distributions are all roughly normal so we summarise them with the median values and standard deviations in Table~\ref{HyperParameters}. In particular, the two regions are found to have relatively similar distributions in \rr\, but the West-End has a larger average \av\ as well as a larger range in this parameter. Translating the median values in Table~\ref{HyperParameters} into the conventional units, B5 has a median \rv\ of 2.6, and a median \av\ of 2.2 magnitudes while the West-End has a median \rv\ of 2.8 and a median \av\ of 4.3 magnitudes. The low median in B5 is easily seen in Figure~\ref{Positions}, which shows the spatial distribution of inferred \av\ values for both regions. Our solutions for \av\ exhibits spatial structure which is consistent with known cloud features, as seen in Figure~\ref{Positions}.

\begin{figure}
\includegraphics[width=9cm]{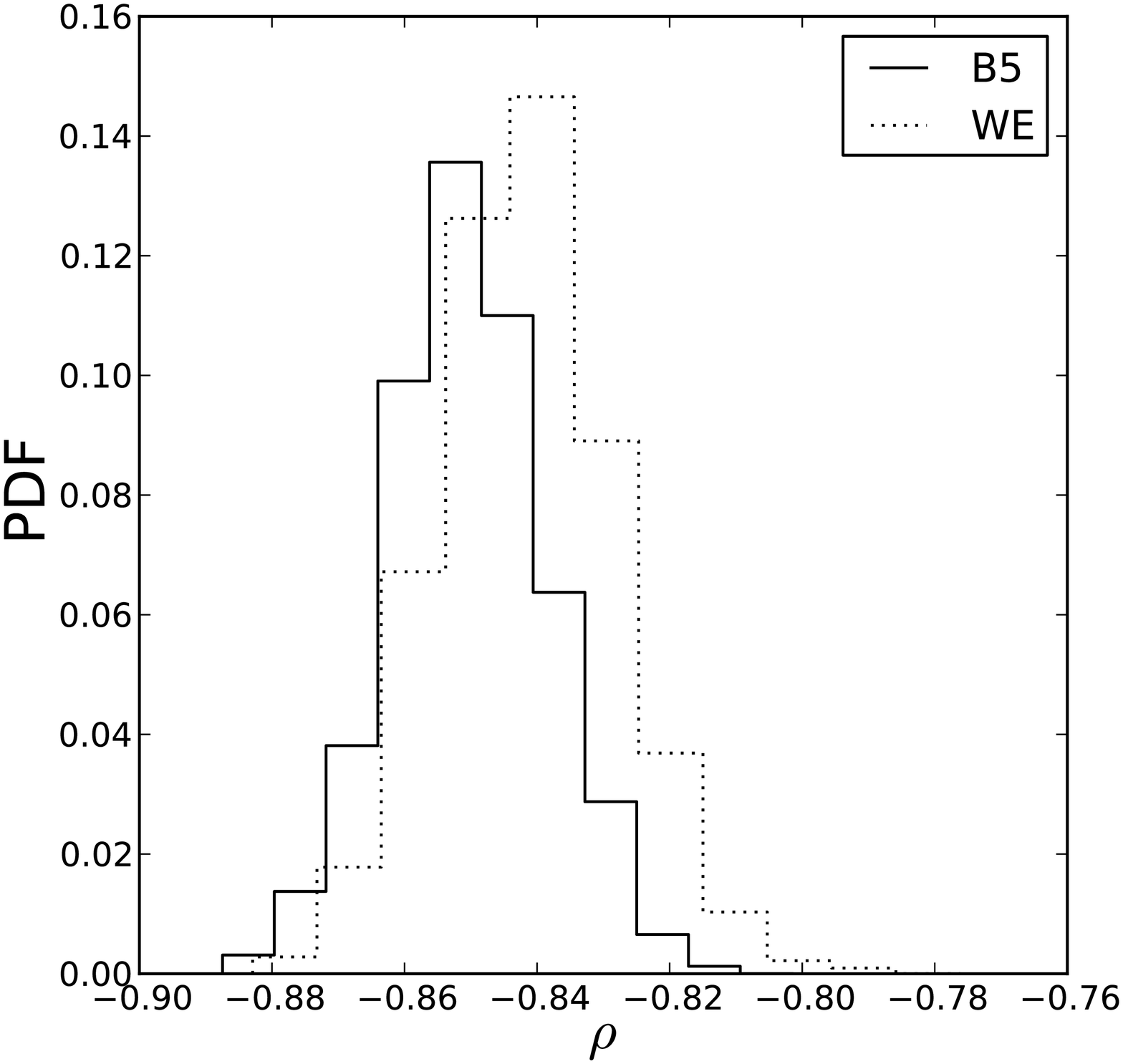}
\caption{The marginal posterior distributions for the hyper-parameter $\rho$, which describes the correlation between \rr\ and $\log(\mav)$, for both B5 (solid) and the West-End (dashed). The two distributions are consistent with each other and inconsistent with $\rho$ = 0. The sense of the correlation is that \rv\ increases with increasing \av.}
\label{Hypers-rho}
\end{figure}

Consistent with the appearance of Figure~\ref{AvRv}, the posterior distribution of $\rho$ is negative and inconsistent with zero. Remember that $\rho$ is the correlation between \rr\ and $\log(A_V)$, and so a negative $\rho$ implies a positive correlation between \av\ and \rv. The posterior distributions of $\rho$ are shown in Figure~\ref{Hypers-rho} and the results from the two regions are consistent with each other. 

The consistency in posterior distributions exists despite the differences between the two regions. In their consideration of sub-regions within Perseus, \citet{Pineda:2008} found that B5 was the most quiescent region with a simple density structure and a simple relationship between column density and CO intensity, while the West-End had a more complicated relationship between column density and CO intensity, suggesting significant clumping within the region. Later studies have found that the quiescent core within B5 \citep{Pineda:2010} is dominated by a very narrow (5000 AU wide) filament \citep{Pineda:2011}. 

\citet{Chapman:2009b} proposed that outflows within two isolated molecular cores may be the cause of regions where the mid-infrared extinction law is inconsistent with dust models as the outflows impact the dust population. As shown in Figure~\ref{B5-picture}, B5 is dominated by a single dense structure, at the heart of which lies B5-IRS1, which drives a powerful multi-parsec molecular outflow \citep{Bally:1996, Yu:1999, Arce:2010}. Most of the high-extinction stars in the B5 data-set come from areas near this outflow.  The West-End hosts a few impressive outflows in L1448, but also a large number of isolated, dense, cores with little or no star formation (see Figure~\ref{WE-picture}). L1448 is mostly outside the UKIDSS coverage of this region. Thus the majority of the West-End stars at high extinction are found in cores without strong outflows.

Geometrical differences between the two regions could also be expected to produce different correlations between \av\ and \rv. Recall our basic assumption that \rv\ $\propto$ $\avg{a}$ $\propto$ $n$ $\propto$ \av, where $n$ is the volume density and $\avg{a}$ is the average size of dust grains. Increasing the scatter in the connection between $n$ and \av\ would tend to increase the scatter between \rv\ and \av\ and thus decrease any measured correlation. As described above, B5 is a relatively simple large single structure. Velocity information from the COMPLETE CO survey \citep{Ridge:2006} and an exhaustive search for outflows and shells throughout Perseus by \citep{Arce:2010, Arce:2011} suggest that the region is simple in velocity space too. Here, the assumption that column density is simply related to volume density is probably reasonably robust. In the West-End the story is quite different. A large shell appears to be disturbing the entire region, and there are multiple velocity components toward some positions---and therefore likely multiple independent clouds along some lines of sight. These additional velocity components have relatively low antenna temperatures and so do not contribute much to the measured column density along the line of sight, maintaining the connection between column density and volume density. The consistency of our results for these two regions suggests that geometrical effects are not a significant problem. 

We compute a simple estimate of the volume density at which we are seeing significant changes in the extinction law. By a column density of \av\ = 10 magnitudes we have typical \rv\ values of 4-5. \citet{Arce:2011} estimates the width of B5 to be no more than 0.5 pc. Ten magnitudes of visual extinction corresponds to roughly 10$^{22}$ cm$^{-2}$ H$_{2}$ atoms \citep{Bohlin:1978}. If we simply assume a constant volume density we can calculate a typical volume density at 10 magnitudes of \av\ inside B5 is 6.6$\times$10$^{3}$ cm$^{-3}$. Because of the density substructure within B5 \citep{Pineda:2011}, this estimate represents a lower limit on the volume density by which point we are seeing clear evidence of grain growth. 

\section{Conclusion}
\label{sec:conclusion}
We have used $r,i,z$ deep data from Megacam on the MMT combined with UKIDSS $J,H,K$ to probe the extinction law in the column density range 0.1 mag $<$ \av\ $<$ 10 mag using a hierarchical Bayesian model. The Bayesian model allows us to place a well-motivated prior on \av\ (log-normality), and work in a consistent framework with the correlation between \av\ and \rv\ treated as just another parameter for which we compute the posterior distribution. This model is implemented using a generalised Gibbs sampler to generate Monte Carlo Markov Chains which explore the full joint parameter space. The implementation has undergone a series of tests using synthetic data similar to our study data, and performs well. 

We find evidence that the extinction law changes over a relatively narrow range of column densities, rising from an \rv\ $\sim$ 3 at 2 magnitudes of \av\ to an \rv\ $\sim$ 5 by 10 magnitudes of \av. Below 2 magnitudes of \av\ we have insufficient sensitivity to infer a relationship between \rv\ and \av. The two regions in our study, B5 and the West-End lie at opposite ends of Perseus and have different physical conditions, yet both show a strong correlations between \rv\ and \av, suggesting that the steepening of the extinction curve (most likely via grain growth) is a fairly universal process.

We deliberately do not provide an explicit relation predicting the value of \rv\ for a given value of \av. These relations are slightly different for the two different regions and are effectively determined only over a narrow range of \av. In addition, we expect the underlying predictor of \rv\ to be the volume density, $n$, not \av, and the relation between $n$ and any observed \av\ is likely to be highly variable in different regions. This is particularly the case for observations of more distant objects, when substantial column density can be the result of a long path through diffuse matter.

Several recent studies have used Sloan Digital Sky Survey data to study dust reddening and extinction \citep[e.g.][]{Jones:2011, Schlafly:2011, Schlafly:2010}. SDSS data is typically not of sufficient quality in high-extinction regions (we examined the portion of Taurus covered in SDSS-SEGUE) to be useful in constraining the extinction law. Future surveys with deeper coverage in the red-optical (Pan-STARRs and LSST) and broader sky coverage will provide the necessary depth to apply this technique to a large number of molecular clouds without the need for dedicated deep observations. 

\section*{Acknowledgements}
This material is based upon work supported by the National Science Foundation under Grant No. AST-0908159, AST-0407172 and AST-0907903. Funding for SDSS-III has been provided by the Alfred P. Sloan Foundation, the Participating Institutions, the National Science Foundation, and the U.S. Department of Energy Office of Science. The SDSS-III web site is http://www.sdss3.org/. SDSS-III is managed by the Astrophysical Research Consortium for the Participating Institutions of the SDSS-III Collaboration. This publication makes use of data products from the Two Micron All Sky Survey, which is a joint project of the University of Massachusetts and the Infrared Processing and Analysis Center/California Institute of Technology, funded by the National Aeronautics and Space Administration and the National Science Foundation.

\bibliographystyle{mn2e}
\bibliography{stardust}

\appendix 

\section{Hierarchical Bayesian Model for Stellar Colours and Dust}\label{appendix:model}

Hierarchical Bayesian analysis is a framework for probabilistically modelling multiple sources of randomness and uncertainty underlying observed data and unifies inference for both populations and individuals of those populations.  
Statistical inference with hierarchical models provides a principled method of probabilistic de-convolution of physically distinct and random effects that are combined in the observed data.  Probabilistic inference allows for not only the estimation of each separate effect, but also the exploration of the joint uncertainties and trade-offs between the multiple effects.   It enables the estimation of the statistical characteristics of an underlying intrinsic population distribution while accounting for the distortions in the observed distribution caused by estimation error. This correction is called \emph{shrinkage}. This is accomplished by hierarchical models via \emph{partial pooling}, which combines the individual estimates with population information. Similar issues regarding inferring the intrinsic distributions of inferred quantities in the presence of random error have been discussed and specific Bayesian techniques have been applied by \citet{bkelly07,bkelly07b}, \citet*{hogg10}, \citet{loredohendry10}, \citet{Mandel:2009, Mandel:2011}, among others, in other astrophysical contexts. An excellent statistical reference for hierarchical Bayesian modelling is \citet{gelman_bda}.

Our statistical model includes a population distribution that models the intrinsic stellar locus of colours, a population distribution for the the dust extinction and reddening to each star, and incorporates the measurement error for the observed colours for each star.  Using Bayes' Theorem, probabilistic estimates for the unknown parameters of individual stars, as well as the hyper-parameters of the populations, are coherently derived.  In the following sections, we build the components of our statistical model and then derived the global posterior density of the unknown parameters conditional on the data.

\subsection{The Observed Colour Likelihood Function}

Let $\bm{C}$ be a set of linearly independent intrinsic colours of a star.  For example, $\bm{C} = (r-i, i-z, z-J, J-H, H-K_s)$.  This represents the colours of the stars in the absence of dust.  Let $\bm{O}$ be the set of observed colours in the same bands with estimates of the measurement uncertainty described by a covariance matrix $\bm{\Sigma}_O$.  The dust absorption in a particular band $F$ for a given $A_V$ and $r_V \equiv R_V^{-1}$ is modelled using the CCM law: $A_F = A_V ( a_F + b_F r_V)$, where the coefficients $a_F$ and $b_F$ are determined from stellar spectra.  Let $\bm{E}(A_V, r_V) \equiv A_V (\Delta \bm{a} + \Delta\bm{b}\, r_V)$ be a vector of colour excesses in the selected colours for a given $A_V$, $r_V$, and $\Delta \bm{a}$ is a fixed vector with elements, e.g. $(a_r-a_i)$, and $\Delta \bm{b}$ is defined similarly.  Then, conditional on the intrinsic colours and the dust parameters, the observed colours are $\bm{O} = \bm{C} + \bm{E}(A_V, r_V) + \bm{\epsilon}$.  Under the assumption of Gaussian measurement noise, the likelihood function is:
\begin{equation}\label{eqn:obscol_lkhd}
P(\bm{O} | \, \bm{C}, A_V, r_V) = N( \bm{O} | \, \bm{C} + \bm{E}(A_V, r_V) , \bm{\Sigma}_O ).
\end{equation}
The multivariate Gaussian probability density in vector $\bm{y}$ with mean $\bm{\mu}_y$ and covariance matrix $\bm{\Sigma}_y$ is denoted by $N(\bm{y} | \, \bm{\mu}_y , \bm{\Sigma}_y)$.

\subsection{The Intrinsic Colour Population Model}

We can construct a population model for the intrinsic colours of stars using the empirical results of Covey et al. 2007 (C07) based upon analyses of the un-reddened stellar locus using SDSS and 2MASS data.  They use the intrinsic $g-i$ colour as a proxy for spectral class.   We shall refer to this variable as $x = g-i$.  Conditional on $x$, we can construct a normal distribution of colours using the estimated means and ``pseudo-standard deviations'' computed by C07 for each bin in $x$.  Here $x$ ranges from 0.1 to 4.3.  Hence, we can write for any individual star
\begin{equation}\label{eqn:stellarlocus}
P( \bm{C} | \, x ) = N( \bm{C} | \, \bm{\mu}_C(x), \bm{\Sigma}_C(x) )
\end{equation}
where $\bm{\mu}_C(x)$ and $\bm{\Sigma}_C(x)$ are known functions of $x$ based upon the locus values of Table 1 of C07.  Since they do not provide estimates of the colour correlations for a given $x$, we set the diagonal elements of the covariance matrix to the squared pseudo-standard deviations in each colour, and the off-diagonal terms to zero.  To complete the intrinsic population model, we should specify a distribution $P(x)$. We find it is sufficient to take a conservative approach and assume a flat prior on $x$: $P(x) \propto 1$ over the range of $x$.

It will be convenient to analytically integrate out the intrinsic colours $\bm{C}$.   The marginal likelihood of the observed colours for a single star, given its parameters $x$, $A_V$ and $r_V$ is then:
\begin{equation}\label{marginal_lkhd}
\begin{split}
P( \bm{O} | \, x, A_V, r_V) &= \int d\bm{C} \, P(\bm{O} | \, \bm{C}, A_V, r_V) P( \bm{C} | \, x) \\ 
&= N[ \bm{O} | \, \bm{\mu}_C(x) + \bm{E}(A_V, r_V), \bm{\Sigma}_O + \bm{\Sigma}_C(x) ]
\end{split}
\end{equation}

\subsection{Dust Population Model with Correlations}

If the parameters of dust to a set of stars are drawn from a common population, it is advantageous to model that population.  We may expect the dust to stars to have some central $(\log A_V, r_V)$ values with correlated deviations.  A reasonable choice is a  bivariate Gaussian distribution in $(\log A_V, r_V)$.
\begin{equation}
\left(\begin{array}{c} \mrr\ \\ \log(A_V) \end{array}\right)  \sim N\left[\left(\begin{array}{c} \mu_r \\ \mu_A \end{array} \right), \left(\begin{array}{cc} \sigma^2_r & \rho\ \sigma_A \sigma_r \\ \rho\  \sigma_A \sigma_r & \sigma^2_A \end{array} \right)\right].
\end{equation}
This implies lognormal marginal distribution in $A_V$,
\begin{equation}
P(A_V |\, \mu_A, \sigma_A^2) = N( \log A_V | \, \mu_A, \sigma_A^2) \times A_V^{-1},
\end{equation}
and a marginal distribution $r_V \sim N(\mu_r, \sigma_r^2)$.   The population mean and variance of $\log A_V$ are $\mu_A, \sigma_A^2$, respectively.   The population mean and variance of $r_V \equiv R_V^{-1}$ are $\mu_r, \sigma_r^2$, respectively.  The linear correlation between $\log A_V$ and $r_V$ is $\rho$.  We additionally limit $R_V$ to lie in the physically plausible range between 2 and 5.5 ($0.18 < r_V < 0.5$). 

The above model can also be understood as a Gaussian distribution on $r_V$ coupled with a mean linear regression of $\log A_V$ on $r_V$, i.e. $r_V \sim N(\mu_r, \sigma_r^2)$ and
\begin{equation}\label{eqn:linearA_V_on_r_V}
\log A_V | \, r_V  = \alpha_0 + \alpha_1 (r_V - 0.32)/0.04 + \epsilon
\end{equation}
where the regression coefficients and residual variance are defined by the hyper-parameters.
\begin{eqnarray}\label{eqn:conversions}
\alpha_0 \equiv \mu_A + \rho \frac{\sigma_A}{\sigma_r}(0.32 -\mu_r), & \alpha_1 \equiv 0.04 \rho\frac{\sigma_A}{\sigma_r} 
\end{eqnarray}
The error term $\epsilon \sim N(0, \sigma^2)$ has variance $\text{Var}[\epsilon] \equiv \sigma^2 = \sigma_A^2 (1-\rho^2)$.
Using these equations we can work with either $\{\mu_A, \sigma^2_A, \rho, \mu_r, \sigma_r^2 \}$ or $\{ \alpha_0, \alpha_1, \sigma^2, \mu_r, \sigma_r^2  \}$ and translate between the parameterizations.  For implementing the Gibbs sampler, we choose the latter parameterization, since it is more easily extended to non-linear relations by adding polynomial terms to Eq. \ref{eqn:linearA_V_on_r_V}.

\subsection{The Hierarchical Posterior Probability Density}

To complete the full probability model we must specify the hyper-prior density on the hyper-parameters $\bm{\alpha}, \sigma^2, \mu_r$, and $\sigma_r^2$.  We choose standard diffuse non-informative distributions: uniform distributions on $\mu_r$, $\bm{\alpha}$, $\log \sigma_r^2$ and $\log \sigma^2$.

We can now construct the hierarchical posterior density.  Suppose we have $N_s$ stars with observed colours $\{ \bm{O}_s \}$, $s = 1 \ldots N_s$.  The unknown parameters for each star are the intrinsic colours $\bm{C}_s$, the spectral type $x_s$, and the dust parameters $A_V^s, r_V^s$.   We can, however, analytically integrate out $\bm{C}_s$ and use the marginal likelihood, Eq. \ref{marginal_lkhd}.  Furthermore, there are several hyper-parameters describing the dust probability model: $\bm{\alpha}$, $\sigma^2$, $\mu_r$ and $\sigma_r^2$.  For star $s$, the posterior distribution over all remaining parameters conditioning on the observed colour data and on the population hyper-parameters is 
\begin{equation}
\begin{split}
P( A_V^s, r_V^s, x_s &| \, \bm{\alpha}, \sigma^2, \mu_r, \sigma_r^2; \bm{O}_s)  \\ 
&\propto P( \bm{O}_s | \,  x_s, A_V^s, r_V^s) \times P(x_s) \\
& \;\;\;\; \times P(A_V^s, r_V^s | \,\bm{\alpha}, \sigma^2, \mu_r, \sigma_r^2) \\
\end{split}
\end{equation}
Since the hyper-parameters are unknown and must be estimated jointly with the parameters from the data, we must compute the joint posterior density over all unknown parameters and hyper-parameters conditioned on the entire data set.  This is just a product of $N$ terms for each star times the hyper-prior.
\begin{equation}\label{posterior}
\begin{split}
P( &\{ A_V^s, r_V^s, x_s\} ; \bm{\alpha}, \sigma^2, \mu_r, \sigma_r^2 | \, \{\bm{O}_s\}) \propto\\
& \left[ \prod_{s=1}^{N} P(  A_V^s, r_V^s, x_s | \, \bm{\alpha}, \sigma^2, \mu_r, \sigma_r^2; \bm{O}_s) \right] \times P(\bm{\alpha}, \sigma^2, \mu_r, \sigma_r^2)
\end{split}
\end{equation}
All fully Bayesian inferences are based upon the computation of this posterior probability distribution.

\section{MCMC Simulation of the Posterior Probability Density with Gibbs Sampling}\label{appendix:mcmc}

For a data set with $N \sim 2000$ stars, there are $3N+5$ parameters and hyper-parameters to be estimated jointly from the hierarchical posterior distribution, after analytically marginalising out the intrinsic colours $\{ \bm{C}_s \}$. To do this efficiently, we have developed a Markov Chain Monte Carlo algorithm based on Gibbs sampling.

Markov Chain Monte Carlo (MCMC) is a class of algorithms that generate random draws from an arbitrary probability distribution by simulating a stochastic process or random walk through parameter space.  Gibbs sampling is a particular MCMC strategy that uses the information in the set of conditional distributions to make the random moves in the stochastic process. One parameter at a time is updated from its conditional probability density, holding the other parameters fixed at their current values. Cycling this process through all the parameters repeatedly generates a Markov chain that explores parameter space and converges to the joint posterior probability density. Once the random samples are generated, inferences can be computed using these samples.  Further theory and techniques of MCMC can be found in various textbooks, e.g. \citet{liu02, gelman_bda, robert05}.

An advantage of using a Gibbs sampling approach by sequential sampling of the conditional densities compared to a standard Metropolis strategy is that the Gibbs sampling approach does not require tuning the jump size of the proposal distribution.  This is important when probing a high-dimensional parameter space as we do here;  it is practically infeasible to optimise the jump size for thousands of parameters.

A disadvantage of traditional Gibbs sampling is that it allows only for orthogonal moves in the parameter space, as one parameter is sampled conditional on the fixed current values of the other parameters.  This problem can be acute if there are strong posterior degeneracies or correlations between parameters.  For example, in this work, we expect there to be trade-offs between the intrinsic colours and dust reddening in the fit for a single star's observed colours.  This trade-off will depend upon the shape of the intrinsic stellar locus as well as the current estimate of the reddening law parameter $R_V$ for the dust to the star.  If there exist multiple local posterior maxima in parameter space separated along a directions oblique to the parameter axes, the orthogonal nature of the Gibbs sampling moves could cause the Markov Chain to get stuck at a sub-optimal solution.

To alleviate these problems, we have included \emph{generalised} conditional sampling steps to our MCMC algorithm.   These steps enable the chain to move along expected degeneracies between parameters that may be oblique with respect to the natural co-ordinate system defined by the chosen parameters \citep{liu&sabatti00, liu02}.  
This strategy allows for non-orthogonal moves through parameter space that change several parameters at once.  For example, we found it advantageous when fitting for each star to simultaneously reduce dust extinction and increase the intrinsic colour while adjusting the reddening law.   This corresponds to an oblique translation through parameter space described by $A_V \rightarrow A_V + \gamma$, $x \rightarrow x + c_x \gamma$, and $r_V \rightarrow r_V + c_r \gamma$.  The coefficients $(c_r, c_x)$, determining the direction of the translation, are randomly chosen according to pre-determined probability distributions, and the magnitude of the translation $\gamma$ is sampled from the posterior density, such that the stationary distribution of the chain is left invariant.  These generalised conditional sampling steps are described below in Step 4.   Note that the Gibbs sampler without these steps (i.e. using only Steps 1-3, 5-7) generates a valid Markov Chain.  The additional Step 4 is added to speed convergence of the chain and help it escape local maxima.

We start with initial guesses of the unknown parameters and hyper-parameters.  The current position of the Markov Chain is $\mathcal{S} = (\{A_V^s, r_V^s, x_s\} ; \bm{\alpha}, \sigma^2, \mu_r, \sigma_r^2)$.   Our Gibbs sampling strategy proceeds as follows.  At each step we sample a new value of a parameter conditional on all the others, and replace the new value in the state vector $\mathcal{S}$. The first four steps are repeated for all stars, conditional on the current values of the population hyper-parameters.

\begin{enumerate}

\item Draw a new $x_s$ from the conditional posterior density $P( x_s | \cdot, \{\bm{O}_s\}) \propto P( \bm{O}_s | \,  x_s, A_V^s, r_V^s) \times P(x_s)$.  (We use $(\cdot)$ to denote all other parameters and hyper-parameters not written explicitly).  This is done by evaluating Eq. \ref{marginal_lkhd}) on a fine grid in $x_s$ and using the inverse-cdf sampling method. 

\item Draw a new $A_V^s$ from the conditional posterior 
\begin{equation}
\begin{split}
P(A_V^s | \cdot, \{\bm{O}_s\}) &= P(A_V^s | x_s, r_V^s, \bm{\alpha}, \sigma^2; \bm{O}_s) \\ 
&\propto N(A_V^s | \hat{A},  \hat{V}_A ) \times P(A_V^s | r_V^s, \bm{\alpha}, \sigma^2), 
\end{split}
\end{equation}
where $\hat{A} = \hat{V}_A \bm{\beta}^T \bm{\Sigma}(x)^{-1} [\bm{O}_s -\bm{\mu}_C(x_s)]$ is the least-squares estimate of $A_V$ with variance $\hat{V}_A = (\bm{\beta}^T \bm{\Sigma}(x)^{-1} \bm{\beta})^{-1}$, and $\bm{\beta} = \Delta\bm{a} + \Delta\bm{b} \, r_V$ and $\bm{\Sigma}(x) = \bm{\Sigma}_O^s + \bm{\Sigma}_C(x_s)$.  The last term is given by Eq. \ref{eqn:linearA_V_on_r_V}.  Since this is non-Gaussian, we obtain a draw by evaluating this on a fine grid in $A_V^s$ and using the inverse cdf sampling method.

\item Sample a new $r_V^s$ from the conditional posterior 
\begin{equation}
\begin{split}
P(r_V^s | \cdot, \{\bm{O}_s\}) &= P(r_V^s | x_s, A_V^s, \bm{\alpha}, \sigma^2, \mu_r, \sigma_r^2, \bm{O}_s)  \\
 &\propto N(r_V^s | \hat{r}_V, \hat{\sigma}_r^2) \times N( r_V^s | \mu_r, \sigma_r^2) \\
 & \;\;\;\; \times P( A_V^s | r_V^s, \bm{\alpha}, \sigma^2 )
\end{split}
\end{equation}
where $\hat{r}_V = \hat{\sigma}_r^2 A_V^s \Delta \bm{b}^T \bm{\Sigma}_O^{-1} ( \bm{O}_s -\bm{\mu}_C(x_s) -A_V^s \Delta \bm{a})$ is the least-squares estimate of $r_V$ and $\hat{\sigma}_r^2 = (\Delta \bm{b}^T \Sigma_O^{-1} \Delta \bm{b} (A_V^s)^2 )^{-1}$ is its variance.  A new $r_V^s$ is generated by evaluating this on a fine grid in $r_V^s$ between 0.18 and 0.50 and using the inverse-cdf sampling method.

\item Generalised Gibbs sampling.  We translate along directions involving changes in the three parameters:
\begin{eqnarray}\label{eqn:generalised}
A^s_V \rightarrow A^s_V + \gamma, & x_s \rightarrow x_s + c_x \gamma, & r^s_V \rightarrow r^s_V + c_r \gamma.
\end{eqnarray}
The direction is randomly chosen each time from the following distributions:
\begin{equation}
c_x \sim \frac{1}{3} N(-0.60, 0.05^2) + \frac{1}{3} N(-0.55, 0.02^2) + \frac{1}{3} N(-0.50, 0.05^2)
\end{equation}
\begin{equation}
c_r \sim \frac{1}{3} N(0.03, 0.02^2) + \frac{1}{3} N(0.05, 0.02^2) + \frac{1}{3} N(0.08, 0.02^2),
\end{equation}
and the sign of $c_r$ is flipped with 50\% probability.  These distributions were chosen via experimentation and were found to significantly help the mixing of the chains.  The shift $\gamma$ is sampled from
\begin{equation}
P(\gamma) \propto P( A^s_V + \gamma , r^s_V + c_r \gamma, x_s + c_x \gamma| \, \bm{\alpha}, \sigma^2, \mu_r, \sigma_r^2; \bm{O}_s) 
\end{equation}
by evaluating this density on a fine grid in $\gamma$ and using the inverse-cdf method.  Then with $\gamma, c_x, c_r$ chosen, the translation Eqs. \ref{eqn:generalised} is performed.

\item Steps 1-4 are repeated for all stars.  Once all individual parameters have been updated, we update the hyper-parameters.  First, we sample from the joint conditional posterior density $P(\mu_r, \sigma_r^2 | \cdot, \{\bm{O}_s\}) = P(\mu_r |  \bar{r}_V, \sigma_r^2) P(\sigma_r^2 | s_r^2)$, where $\bar{r}_V$ is the sample mean of the current $r_V^s$ of all stars, and $s^2_r$ is the sample variance.  This is done by drawing $\sigma_r^2$ from a scaled inverse chi-squared distribution and, conditional on that, drawing $\mu_r$ from a normal:
\begin{equation}
\sigma_r^2 | \, s^2_r\sim \text{Inv-}\chi^2(N -1, s_r^2)
\end{equation}
\begin{equation}
\mu_r | \, \bar{r}_V, \sigma^2_r \sim N(\bar{r}_V, \sigma_r^2 /N).
\end{equation}

\item Next we sample from the joint conditional posterior $P( \bm{\alpha}, \sigma^2 | \cdot, \{\bm{O}_s\}) = P( \bm{\alpha}, \sigma^2 | \{ \log A_V^s, r_V^s\})$.  This is equivalent to the Bayesian ordinary linear regression problem.  If we let $\bm{Y}$ be vector with elements $\log A_V^s$, then we have $\bm{Y} | \bm{\alpha}, \sigma^2, \{r_V^s\} \sim N( \bm{D} \bm{\alpha}, \bm{I} \sigma^2)$.  The design matrix $\bm{D}$ has rows $\bm{D}_s = [1, (r_V^s-0.32)/0.04]$.  Compute $\hat{\bm{V}}_\alpha = (\bm{D}^T \bm{D})^{-1}$, $\hat{\bm{\alpha}} = \hat{\bm{V}}_\alpha \bm{D}^T \bm{Y}$, and
\begin{equation}
S^2_Y = \frac{1}{ (N-2)} (\bm{Y} - \bm{D} \hat{\bm{\alpha}})^T  (\bm{Y} - \bm{D} \hat{\bm{\alpha}}).
\end{equation}
Gibbs sampling proceeds by drawing a new $\sigma^2$ from a scaled inverse chi squared distribution, and then, conditional on that, sampling new $\bm{\alpha}$ from a multivariate normal:
\begin{equation}
\sigma^2 | \, S^2_Y \sim \text{Inv-}\chi^2(N-2 |\, S^2_Y)
\end{equation}
\begin{equation}
\bm{\alpha} | \, \bm{Y}, \bm{D}, \sigma^2 \sim N( \hat{\bm{\alpha}}, \hat{\bm{V}}_\alpha \sigma^2)
\end{equation}

\item At this point, we have updated every parameter and hyper-parameter, and we record the state of the chain $\mathcal{S}$.  We return to step 1 and iterate the Gibbs sampler with the current parameter values many times until convergence.  

\end{enumerate}

We typically run multiple independent chains in parallel starting from randomised positions and monitor convergence using the Gelman-Rubin ratio \citep{gelman92}.

\label{lastpage}

\end{document}